\documentclass{epl2} 
\usepackage{amssymb,cite,amsfonts,amsmath,graphicx,amsmath,amsbsy,mathrsfs,subfigure,float}
\usepackage{pifont,geometry,txfonts,multicol}
 \usepackage[titletoc]{appendix}
\renewcommand{\figurename}{Fig.}
\newcommand{\ucite}[1]{\textsuperscript{\cite{#1}}}
\newcommand{\kref}[1]{Eq.\ $\left(\ref{#1}\right)$}
\newcommand{\bref}[1]{$\left(\ref{#1}\right)$}

\renewcommand{\onecolumn}[1]{\end{multicols}  {#1} \begin{multicols}{2}}
\begin{document}
	\title{Dynamic Entanglement Evolution of Two-qubit XYZ Spin Chain in Markovian Environment}
	\author{Yi-Chong, Ren\inst{1}\thanks{Email: rych@mail.ustc.edu.cn} \and Hong-Yi, Fan\inst{1}}
	\institute{
		\inst{1} Department of Material Science and Engineering, University of Science
		and Technology of China, Hefei city, Anhui prov. , China
	}
	\abstract{
		We propose a new approach called Ket-Bra Entangled State (KBES) Method for converting master equation into Schr\"{o}dinger-like equation. With this method, we investigate decoherence process and entanglement dynamics induced by a $2$-qubit spin chain that each qubit coupled with reservoir. The spin chain is an anisotropy $XYZ$ Heisenberg model in the external magnetic field $B$, the corresponding master equation is solved concisely by KBES method; Furthermore, the effects of anisotropy, temperature, external field and initial state on concurrence dynamics is analyzed in detail for the case that initial state is Extended Wenger-Like(EWL) state. Finally we research the coherence and concurrence of the final state (namely the density operator for time tend to infinite).}

\maketitle
\begin{multicols}{2}
\section{Introduction}

Entanglement and coherence that the most distinguishing features of quantum mechanics play a crucial role in different fundamental aspects of quantum theory and practical applications in quantum-information processing\ucite{r1,r2}; Decoherence process as a consequence of the unavoidable interplay between system and reservoir, through which quantum superposition states are irreversibly decay into classical and statistical mixture states; Along with the loss of coherence, disentanglement will decrease the reliable of quantum communication and computation\ucite{r3} significantly.

In recently years, the aspect that has mostly drawn attention is the research
of entanglement properties of condensed matter systems and their application in
quantum information; The intrinsic characteristic and potential importance of
decoherence and disentanglement in condensed state, lead to a flow of analysis
in solid state especially in spin chains\ucite{r4,r5,r6}
potential applications of entanglement in solid state have simulated research on ways to improve the usefulness time and control of entanglement;
Compared with other physics systems, spin chains are the most competitive candidates for the
realization of quantum computation\ucite{r7,r8,r9,r10}; More and more interest has arisen in the
dynamic evolution of spin chains exposed to local thermal and Markovian noisy
that has applications not only for gate operation in solid state quantum
computation, but also for quantum state transfer in quantum communication.

However, in the literatures most research was focused on thermal entanglement\ucite{r11,r12,r13} and intrinsic decoherence\ucite{r14} of spin chains; There are only a few papers analyzing the dynamic
evolution of spin chains in dephasing channel\ucite{r15}. Therefore we need further
research on spin chains coupled with various environments and noisy.
Recently, a new procedure called Ket-Bra Entangled State(KBES) method has been
proposed that allow one to solve master equation effectively by translating
master equation into Schr\"{o}dinger-like equation, which might be applicable for tackling any
finite level system and any initial state. The objective of present research is to analyze a $2$-qubit anisotropic XYZ Heisenberg chain, each interacts with a local reservoir. Then we shall give
the corresponding master equation and solve it with KBES method. Finally, we
shall use the solution to investigate the entanglement dynamics of the spin
chain under different conditions, while the initially described by a Extended
Werner-Like State(EWLS). This will permit us to find how entanglement
dynamics and its revivals are related to physical parameters like the purity,
and the entanglement of initial state and the temperature, etc..

\section{Super-Operator Method Versus KBES Method}

Here, after briefly reviewing the general super-operator method, we introduce Fan's method which can transform general operator between real and fictitious mode by constructing the bosonic thermal-entangled representation\ucite{s4}; Based on Fan's work, we propose our KBES method which can solve master equation by converting it into Schr\"{o}dinger-like equation.

To expound the super-operator method\ucite{s7}, consider an usual master equation
\begin{equation}
\dot{\rho}=\mathscr{L}\rho, \label{d1}
\end{equation}
Lindblad operator $\mathscr{L}$ is given by
\begin{align}
\mathscr{L}=  &  \frac{\gamma}{2}\left(  n+1\right)  \left(  2\sigma^{-}%
\rho\sigma^{+}-\sigma^{+}\sigma^{-}\rho-\rho\sigma^{+}\sigma^{-}\right)
\nonumber\\
&  +\newline\frac{\gamma}{2}n\left(  2\sigma^{+}\rho\sigma^{-}-\sigma
^{-}\sigma^{+}\rho-\rho\sigma^{-}\sigma^{+}\right)  . \label{d9}%
\end{align}
When $\mathscr{L}$ is independent of time, the form solution of \kref{d1} is
\begin{equation}
\rho\left(  t\right)  =e^{\mathscr{L}t}\rho\left(  0\right)  . \label{d2}%
\end{equation}

$\rho\left(  t\right)  $ can be given in explicit expression only for
$\mathscr{L}$ consist of super-operator generators of some Lie algebras, of course, there is a wide class of master equation satisfy the condition. For \kref{d1} three super-operators are defined by
\begin{equation}
\begin{array}
[c]{l}
L_{+}\rho=\sigma^{+}\rho\sigma^{-},L_{-}\rho=\sigma^{-}\rho\sigma^{+},\\
L_{z}\rho=\frac{1}{2}\left(  \sigma^{+}\sigma^{-}\rho-\rho\sigma^{+}\sigma
^{-}\right)  .
\end{array}
\label{d3}
\end{equation}
Noting that $\left[  L_{+},L_{-}\right]  =2L_{z},\left[  L_{z},L_{\pm}\right]  =\pm L_{\pm
}$, namely $L_{\pm,z}$ is a Su(2) Lie algebra.
Thus $\mathscr{L}=-\left(  n+1/2\right)  \gamma+n\gamma L_{-}+\left(n+1\right)  \gamma L_{+}-\gamma L_{z}.$
With last equation, \kref{d2} can be represented as
\begin{align}
\rho\left(  t\right)   &  =e^{-\frac{1}{2}\left(  2n+1\right)  \gamma
	t+n\gamma L_{-}+\left(  n+1\right)  \gamma tL_{+}-\gamma tK_{z}}\rho\left(
0\right)  ,\nonumber\\
&  =e^{x_{+}\left(  t\right)  L_{+}}e^{\ln x_{z}\left(  t\right)  L_{z}%
}e^{x_{-}\left(  t\right)  L_{-}}\rho\left(  0\right)  , \label{d6}%
\end{align}
where $x_{\pm}\left(  t\right)  ,x_{z}\left(  t\right)  $ have been given in Ref.\ucite{r20}. After transforming the super operator into general operator, the solution of \kref{d6} can be given.

Ulteriorly, prof. Fan introduce the thermal-entangled state
$|I\rangle=e^{a^{\dag}\tilde{a}^{\dag}}|0,\tilde{0}\rangle$ by introducing an extra fictitious mode\ucite{s4}, when bosonic operator acts on the thermal-entangled state
\begin{equation}
a|I\rangle =\tilde{a}^{\dag}|I\rangle, \text{\qquad }a^{\dag}|I\rangle=\tilde{a}|I\rangle. \label{d11}
\end{equation}
The explicit expression of super-operator $L_{\pm,z}$ consists of creation and annihilation operator is
\begin{small}
\begin{equation}%
\begin{array}
[c]{c}%
L_{-}\rho|I\rangle\equiv a\rho a^{\dag}|I\rangle=a\tilde{a}\rho|I\rangle
\Rightarrow L_{-}=a\tilde{a},\\
L_{+}\rho|I\rangle\equiv a^{\dag}\rho a|I\rangle=a^{\dag}\tilde{a}^{\dag}%
\rho|I\rangle\Rightarrow L_{-}=a^{\dag}\tilde{a}^{\dag}
\end{array}
\label{d12}%
\end{equation}
\end{small}
Thermal-entangled state $|I\rangle$ can transform general operator into a fictitious mode, thus it be used to solve master equation of Bose system.

Next, we shall introduce our new method, consider $\rho=
\sum_{m,n}
\rho_{m,n}\left\vert m\right\rangle \left\langle n\right\vert$ in Hilbert space $\mathscr{H}$, $\left\vert m\right\rangle$ constitutes a set of complete orthogonal basis. Then we construct the two-mode entangled state $\left\vert\eta\right\rangle =\sum_{n}\left\vert n,\tilde{n}\right\rangle $, where the tilde\textquotedblleft$\sim$\textquotedblright\ means the extra fictitious mode, then:
\begin{equation}
\left\vert \rho\right\rangle =\rho\left\vert\eta\right\rangle ={\displaystyle\sum\limits_{m,n}}
\rho_{m,n}\left\vert m,\tilde{n}\right\rangle . \label{a2}
\end{equation}
\kref{a2} exhibit that the extra fictitious "$\sim$" mode $\left\vert \tilde{n}\right\rangle $ indeed represent the Bra-vector $\left\langle n\right\vert $, thus we call $\left\vert \eta\right\rangle $ the Ket-Bra Entangled State(KBES).
For any operator$A_{mn}\equiv\left\vert m\right\rangle \left\langle n\right\vert $ in $\mathscr{H}$,
we have
\begin{equation}
A_{mn}\left\vert \eta\right\rangle 
=\left\vert m,\tilde{n}\right\rangle
=\left\vert \tilde{n}\right\rangle \left\langle \tilde{m}\right\vert
\left\vert \eta\right\rangle =\tilde{A}_{mn}^{\dag}\left\vert \eta
\right\rangle . \label{a3}%
\end{equation}
Besides, \kref{a3} valids for any $A\equiv {\sum_{m,n}}\mathfrak{a}_{mn}A_{mn} (\mathfrak{a}_{mn}\\ \text{is real})$. Namely $A\left\vert \eta\right\rangle={A}^{\dag}\left\vert \eta\right\rangle$.

The case of \kref{d1} has been solved by KBES method in our previous work
\ucite{s6}, here consider a general Lindblad equation for the $N$-level system
\begin{equation}
\frac{d\rho}{dt}=-\frac{i}{\hbar}\left[  H,\rho\right]  +\sum_{n,m=1}%
^{N^{2}-1}h_{n,m}\left(  L_{n}\rho L_{m}^{\dagger}-\frac{1}{2}\left\{  L_{m}^{\dag}L_{n},\rho\right\}  \right)  , \label{a5}%
\end{equation}

To solve \kref{a5}, introduce the corresponding KBES
$\left\vert \eta_{L}\right\rangle =\sum_{n=1}^{N}\left\vert n,\tilde
{n}\right\rangle $, with  \kref{a3}, the
Lindblad equation can be convert into Schrodinger-like equation
\begin{equation}
\frac{d}{dt}\left\vert \rho\right\rangle  =   \mathscr{F}_{L}\left\vert \rho\right\rangle . \label{a6}
\end{equation}
where $\mathscr{F}_{L}\equiv   -\frac{i}{\hbar}(
H-\tilde{H})  +\sum_{n,m=1}^{N^{2}-1}h_{n,m}[  L_{n}\tilde{L}
_{m}-(  \tilde{L}_{n}^{\dagger}\tilde{L}_{m}+L_{m}^{\dagger
}L_{n})/2  ] $. We name \kref{a6} Schr\"{o}dinger-like equation because of $\mathscr{F}_{L}$ may be non-Hermitian. \kref{a6} can be solved through different approach

One way is the evolution operator method, when $\mathscr{F}_{L}$ is time-independent, so
\begin{equation}
\left\vert \rho_{t}\right\rangle =e^{\mathscr{F}_{L}t}\left\vert \rho_{0}\right\rangle .
\label{a7}%
\end{equation}
The explicit matrix expression of $\mathscr{F}_{L}$ can be given in Kronecker product space $\mathscr{H}\otimes
\tilde{\mathscr{H}}$ with the matrix expression of $L_{m}$,
Sometimes we can also decompose $e^{\mathscr{F}_{L}t}$
into several exponential form of operators with Lie algebra.

The other way is the stationary state method.
i.e., to calculate the eigenstates and eigenvalues of $\mathscr{F}_{L}$%
\begin{equation}
\mathscr{F}_{L}\left\vert \varphi_{i}\right\rangle =\lambda_{i}\left\vert \varphi
_{i}\right\rangle. \label{a8}%
\end{equation}
Then the solution can be represented as:
\begin{equation}
\left\vert \rho_{t}\right\rangle =
{\displaystyle\sum\limits_{i}}
C_{i}e^{\lambda_{i}t}\left\vert \varphi_{i}\right\rangle, \label{a9}
\end{equation}
where $C_{i}$ is constant which can be determined by initially $\rho\left(0\right)  $ and characters of density operator. Obviously, {\bf all eigenvalues $\mathbf{ \lambda_{i}\leq0}$ and the eigenstate $\mathbf{ \vert\varphi\rangle}$ whose eigenvalue $\mathbf{ \lambda=0}$ corresponding the final state $\mathbf {\rho\left(\infty\right)}$.} Further more, most methods that ware used for Schr\"{o}dinger equation previously can also be used to solve the Schr\"{o}dinger-like equation.

Through super-operator method is concise and applicable for any initial state, it show a narrow applicable range, even slightly changes to master equation may lead to unsolvable effect.
Compared with the super-operator method, KBES method have three merits: {\bf 1. The procedure has a wide applicable range and applicable for any master equation of finite-level systems in theoretical; 2. More concise, and all calculation can be completed by computer;  3. The method can convert master equation into Schr\"{o}dinger-like equation, namely most methods of Schr\"{o}dinger equation can be used to solve master equation.}

\section{The Model and Master Equation}
The Hamiltonian of $2$-qubit Heisenberg XYZ model is\ucite{r16}%
\begin{align}
H_{spin}  & =B\left(  \sigma_{1}^{z}+\sigma_{2}^{z}\right)  +J\left(
\sigma_{1}^{+}\sigma_{2}^{-}+\sigma_{1}^{-}\sigma_{2}^{+}\right)
\\
& +J\Delta\left(  \sigma_{1}^{+}\sigma_{2}^{+}+\sigma_{1}^{-}\sigma_{2}%
^{-}\right)  +J_{z}\sigma_{1}^{z}\sigma_{2}^{z},\label{a11}
\end{align}
where $J,J_{z}$ is the coupling coefficient and parameter $\Delta\left(  -1<\Delta<1\right)  $ measures the anisotropy in XY plane. 

The master equation of the spin system is
\begin{equation}
\frac{d\rho}{dt}=-i\left[  H_{spin},\rho\right]  +\sum_{i=1,2}\mathscr{L}_{i}\rho.
\label{a14}%
\end{equation}
Defining $\alpha=\gamma\left(  n+1\right)  ,\beta=\gamma n$,
\begin{align}
\mathscr{L}_{i}\rho & =\beta\left(  2\sigma_{i}^{+}\rho\sigma_{i}^{-}%
-\sigma_{i}^{-}\sigma_{i}^{+}\rho-\rho\sigma_{i}^{-}\sigma_{i}^{+}\right)
\nonumber\\
& +\alpha\left(  2\sigma_{i}^{-}\rho\sigma_{i}^{+}-\sigma_{i}^{+}\sigma
_{i}^{-}\rho-\rho\sigma_{i}^{+}\sigma_{i}^{-}\right)  .
\end{align}

Similar with KBES for \bref{a5}, we define the KBES$\vert \eta\rangle  =
\sum_{m,n=0,1}
\vert m,n\rangle \otimes\left\vert \tilde{m},\tilde{n}\right\rangle$ correspond to \kref{a14}.
Then the \kref{a14} can be converted to the Schr\"{o}dinger-like equation:
\begin{equation}
\frac{d}{dt}\left\vert \rho\right\rangle =\mathscr{F}\left\vert \rho\right\rangle.
\label{a16}
\end{equation}
where $\mathscr{F}\equiv i(\tilde{H}_{spin}-H_{spin})+
\sum_{i=1,2}[\beta(2\sigma_{i}^{+}\tilde{\sigma}_{i}^{+}-\sigma_{i}^{-}\sigma_{i}^{+}-\tilde{\sigma}_{i}^{-}\tilde{\sigma}_{i}^{+})+\alpha(2\sigma_{i}^{-}\tilde{\sigma}_{i}^{-}-\sigma_{i}^{+}\sigma_{i}^{-}-\tilde{\sigma}_{i}^{+}\tilde{\sigma}_{i}^{-})]$.

Thus the solution $\left\vert \rho_{t}\right\rangle =e^{\mathscr{F}t}\left\vert \rho_{0}\right\rangle$ can be obtain with the matrix expression of $\mathscr{F}$ on the basis $\left\vert 0\right\rangle _{i}=(0,1)^{T},\left\vert1\right\rangle _{i}=(0,1)^{T}$. The general solution has been obtained with the help of Mathematica 9.0 and computer, however it's is too intricate to present, while the solution whose initial state is EWLS shall be given in next section.
\section{Entanglement Evolution for EWLS}
As is known to all, Bell-like and Werner-like states are significant in both
quantum information and computation process; In this section, we consider the
Extended Werner-like States (EWLS) as initial states, which can reduce to Werner-like states and Bell-like states respectively; Accordingly, we consider the initial states
\begin{equation}%
\begin{array}
[c]{l}%
\rho^{\Phi}\left(  0\right)  =r\left\vert \Phi\right\rangle \left\langle
\Phi\right\vert +\frac{1-r}{4}I_{4},\\
\rho^{\Psi}\left(  0\right)  =r\left\vert \Psi\right\rangle \left\langle
\Psi\right\vert +\frac{1-r}{4}I_{4},
\end{array}
\label{a18}
\end{equation}
$r$ implies the purity of initial states, $\left\vert
\Phi\right\rangle $ and $\left\vert \Psi\right\rangle $ are bell-like states
\begin{equation}
\left\vert \Phi\right\rangle =a\left\vert 01\right\rangle +b\left\vert
10\right\rangle ,\text{ }\left\vert \Psi\right\rangle =a\left\vert
00\right\rangle +b\left\vert 11\right\rangle , \label{a19}%
\end{equation}
where $a$ is real, $b=\left\vert b\right\vert e^{i\delta},a^{2}
+\left\vert b\right\vert ^{2}=1$. EWL states reduce to Bell-like states for
$r=1$ and Winger-like states for $\left\vert a\right\vert =\left\vert
b\right\vert =1/\sqrt{2}$ respectively.

Wootters' concurrence can describe entanglement dynamics, and conveniently for bipartite system.
EWL states belong to "$X$" states,  while concurrence for "$X$" states can be easily calculated because the "$X$" structure is maintained during evolution, that is\ucite{s1,s2}
\begin{equation}
C_{\rho}\left(  t\right)  =2\max\left\{  0,K_{1}\left(  t\right)
,K_{2}\left(  t\right)  \right\}  , \label{a22}%
\end{equation}
Where $K_{1}\left(  t\right)  =\left\vert \rho_{14}\left(  t\right)
\right\vert -\sqrt{\rho_{22}\left(  t\right)  \rho_{33}\left(  t\right)  }$
and $K_{2}\left(  t\right)  =\left\vert \rho_{23}\left(  t\right)  \right\vert
-\sqrt{\rho_{11}\left(  t\right)  \rho_{44}\left(  t\right)  }.$
See the explicit expression of $C_{\rho}\left(  t\right)$ in appendix.


The explicit formulation of $C_{\rho}^{\Phi
}\left(  t\right)  $ and $C_{\rho}^{\Psi}\left(  t\right)  $ shall be given in appendix by substituting \kref{s1} into \kref{a22}.

\section{Intensive Analysis of Concurrence}

\begin{figure}[H]
\setcounter{subfigure}{0} \centering
\subfigure[$a=b=1/\sqrt{2}.$]{
\includegraphics[width=0.22\textwidth]{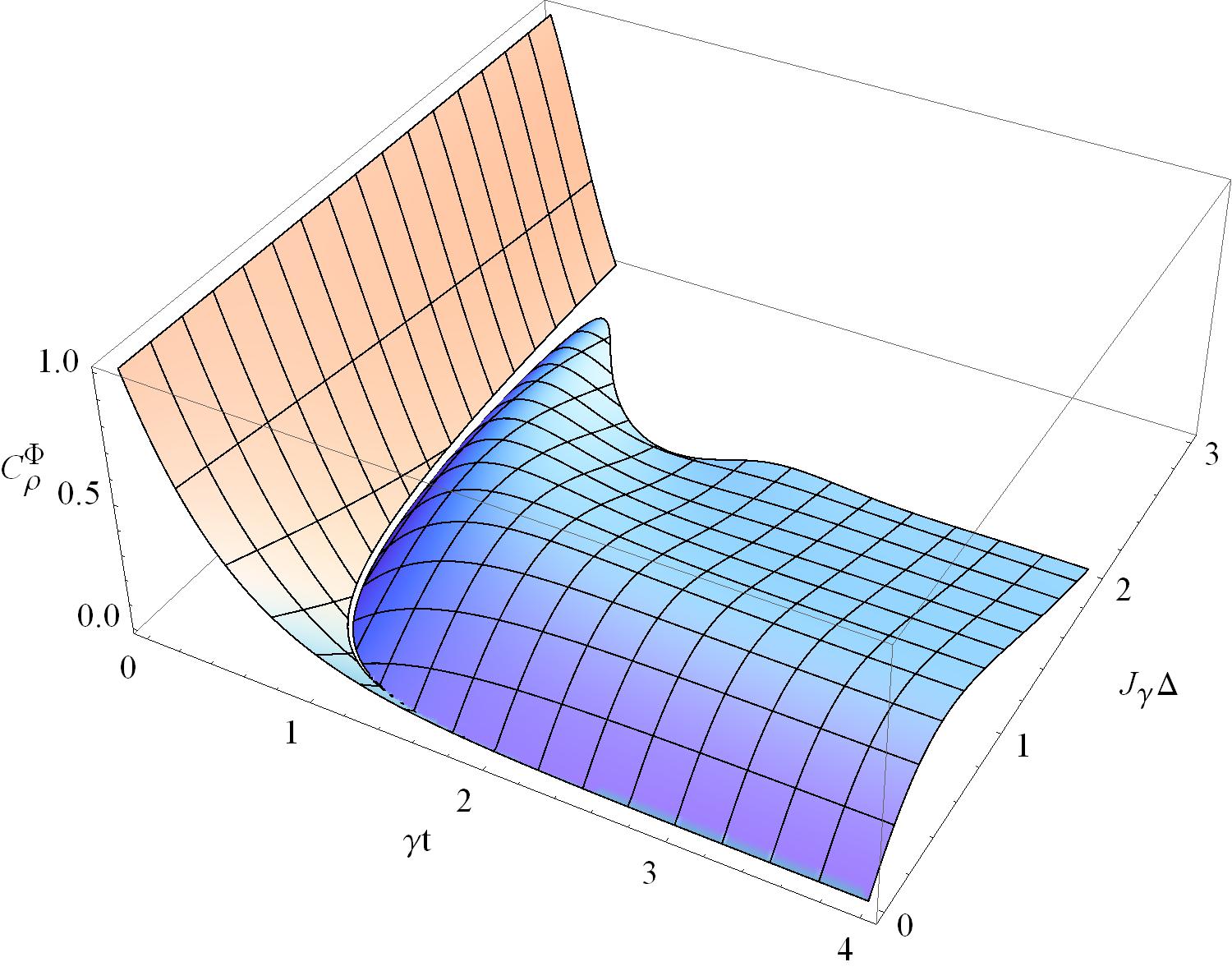}} \hspace{0.05in}
\subfigure[$a=b=1/\sqrt{2}.$]{
\includegraphics[width=0.22\textwidth]{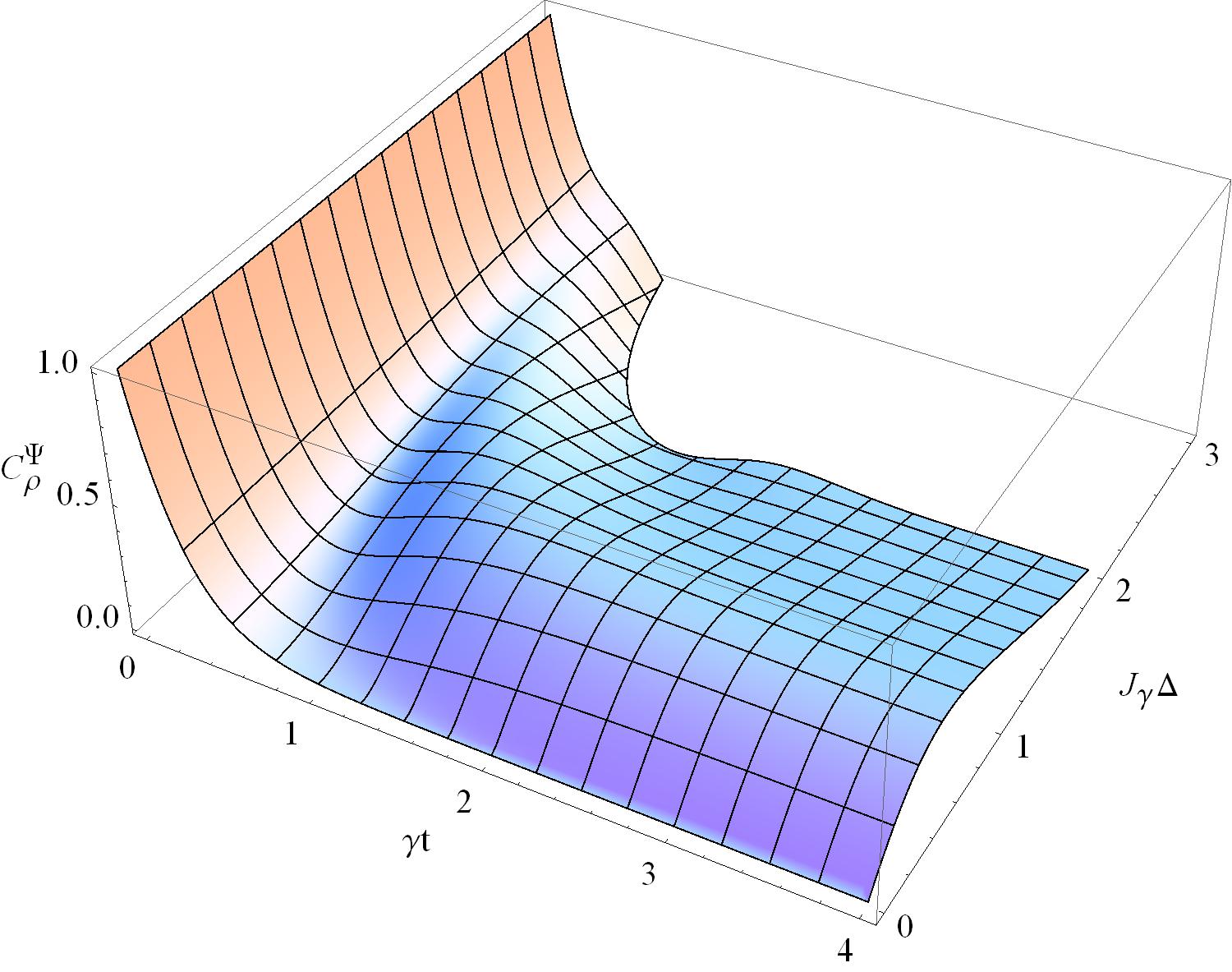}} \\
\subfigure[$a=1,b=0.$]{
\includegraphics[width=0.22\textwidth]{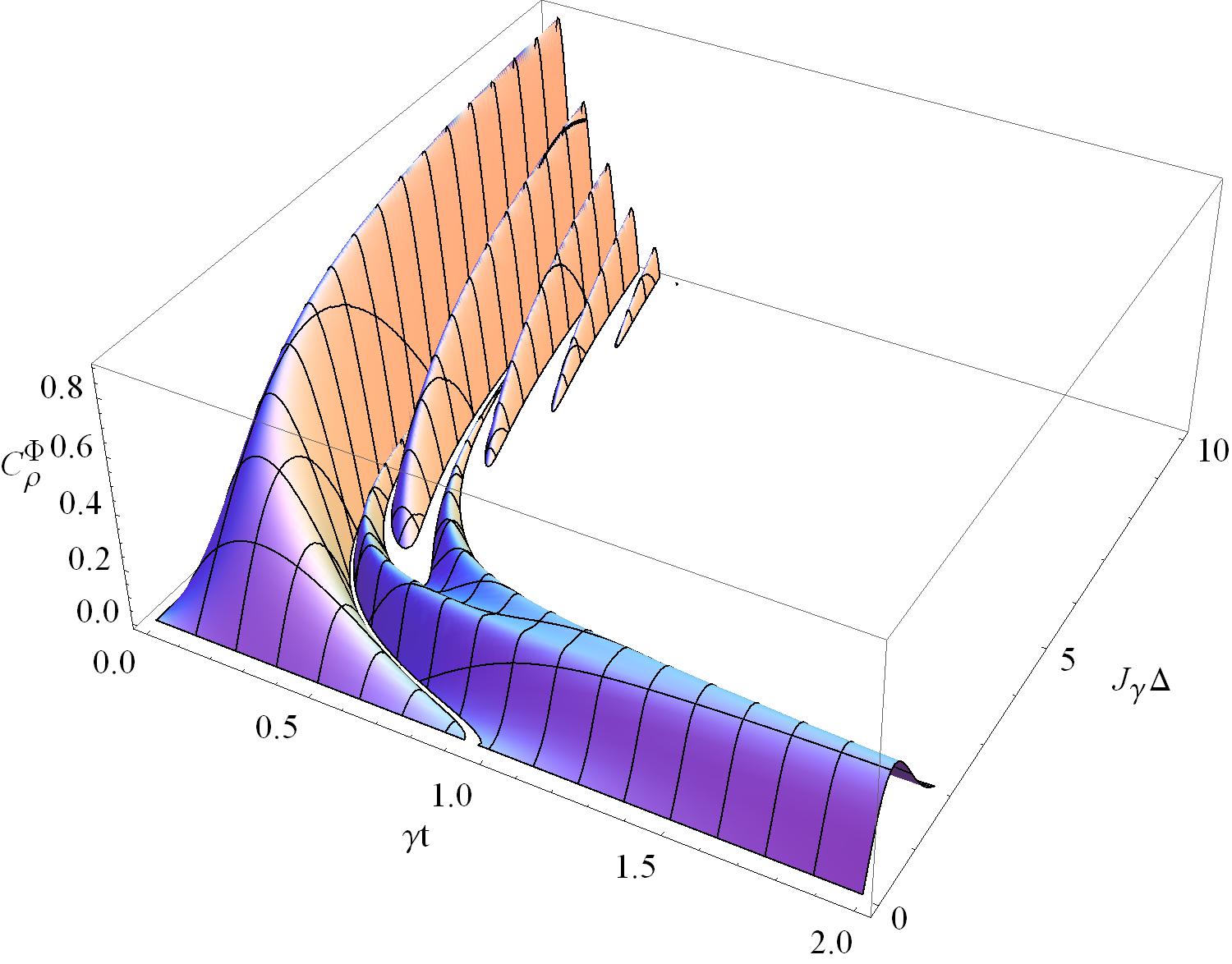}} \hspace{0.05in}
\subfigure[$a=1,b=0.$]{
\includegraphics[width=0.22\textwidth]{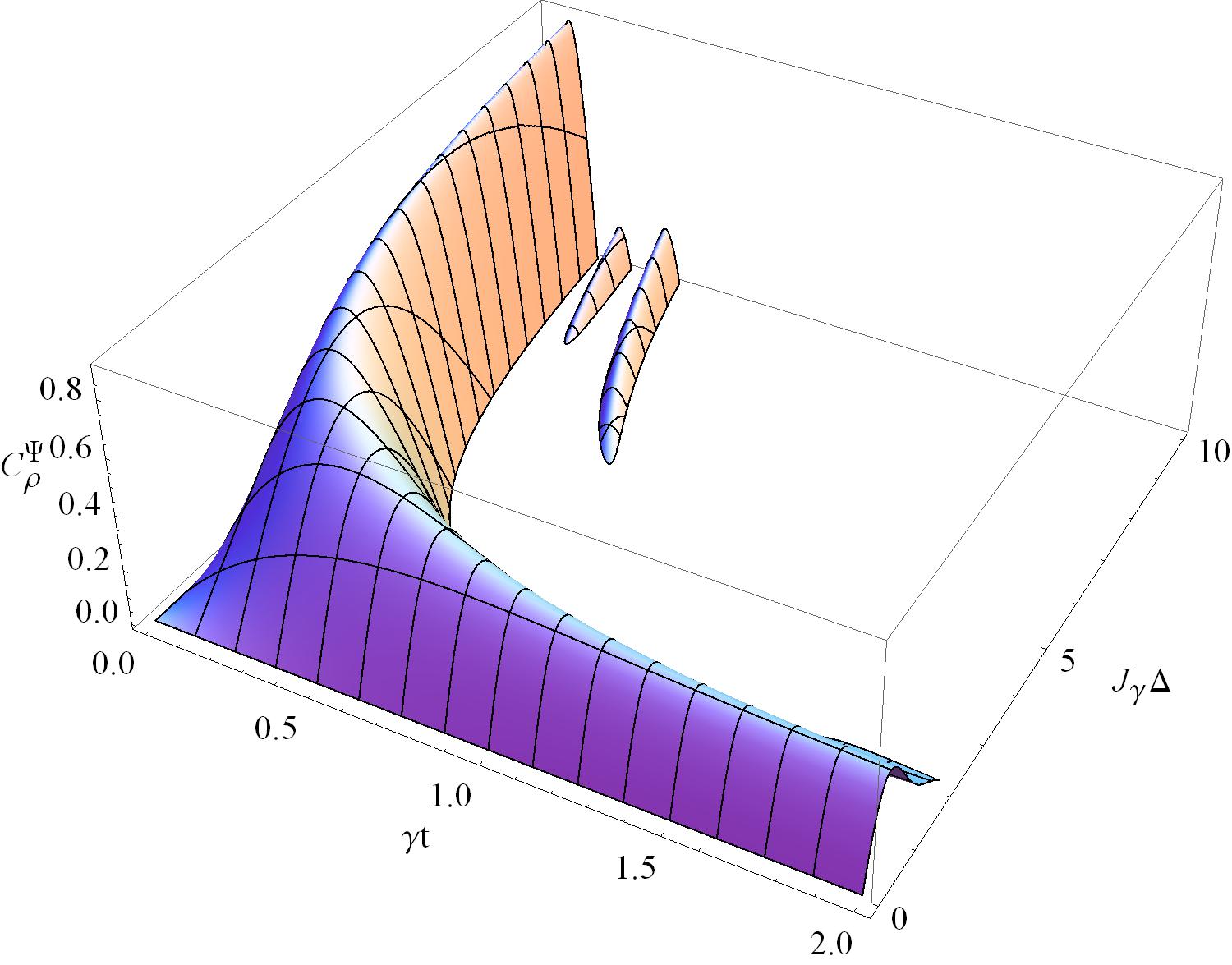}}
\caption{ $C_{\rho}\left(
t\right)  $ versus $\gamma t$ and $J_{\gamma}\Delta$ with $B_{\gamma}=0,r=1,n=0$.}
\label{fig1}
\end{figure}

Eq.\ (\ref{s1},\ref{s2},\ref{a23}) shall be used to analyze the spin chain entanglement dynamics; These equations also display that both $C_{\rho}
^{\Phi}\left(  t\right),C_{\rho}^{\Psi}\left(  t\right)  $ are
independent of $J_{z}$, For $r=1$and $C_{\rho}^{\Psi}$ of \kref{a24}
are plotted in Fig.\ref{fig1}

Fig.\ref{fig1} show that $C_{\rho}^{\Phi}\left(  t\right)  $ consist of two parts $K_{1}^{\Phi}\left(  t\right)  $ and $K_{2}^{\Phi}\left(  t\right)  $, as description in \kref{a23}. For any $C_{\rho}\left(  0\right)  <1$,
both $C_{\rho}^{\Phi}\left(  t\right)  $ and $C_{\rho}^{\Psi}\left(  t\right)
$periodically vanishes with a damping of their revival amplitude; ESD and revival occur for both states but with different frequency, which increases with the growth of $J\Delta$ and $J$. The concurrence no more alway decay to zero 
after sufficient time, this totally differ with the uncoupled case\ucite{s4}, in contrast both $C_{\rho}^{\Phi}\left(  t\right)  $ and $C_{\rho}^{\Psi}\left(
t\right)  $  decay exponentially but vanishes asymptotically and suddenly respectively; 
\begin{figure}[H]
\setcounter{subfigure}{0} \centering
\subfigure[$a=b=1/\sqrt{2}.$]{
\includegraphics[width=0.22\textwidth]{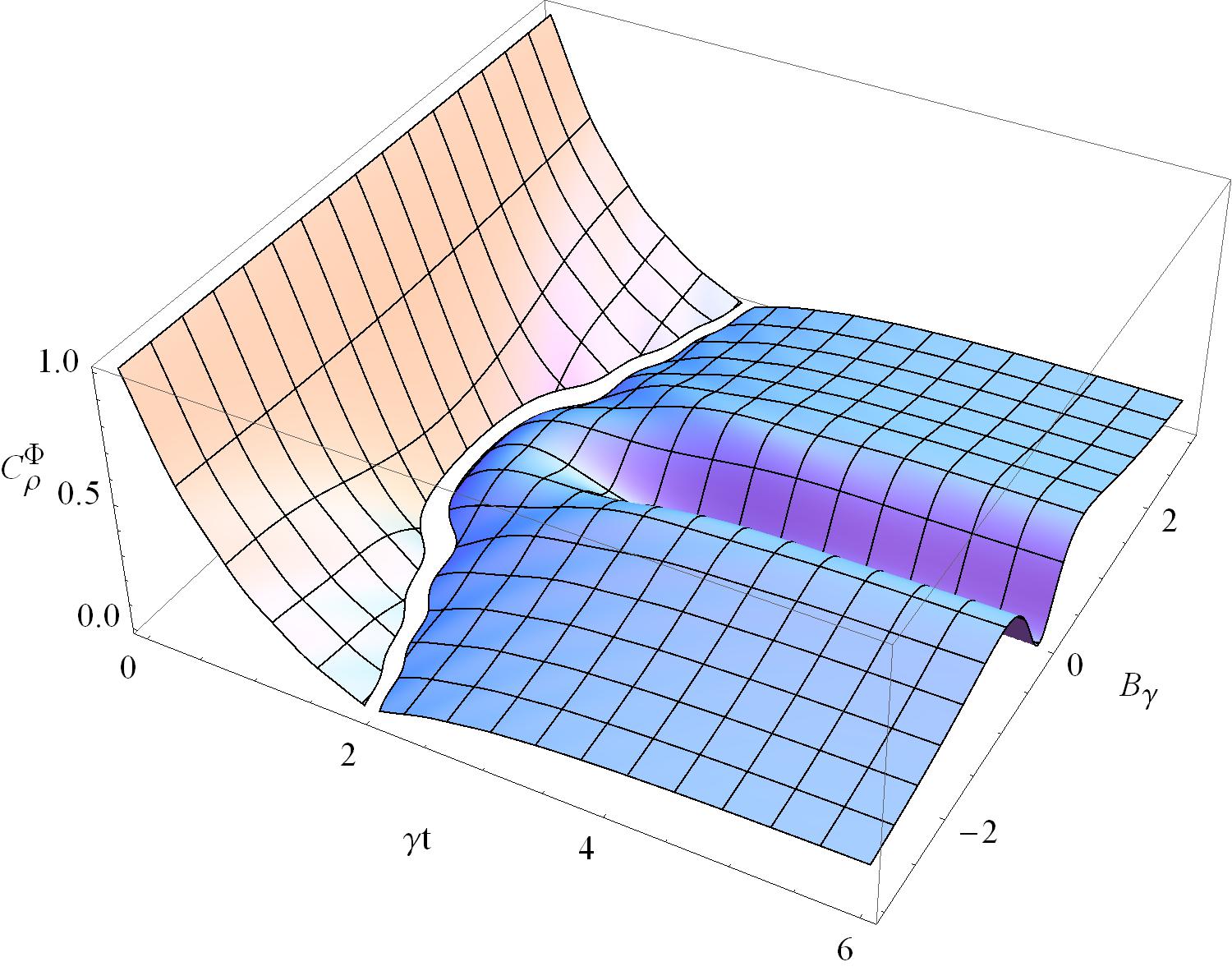}} \hspace{0.05in}
\subfigure[$a=b=1/\sqrt{2}.$]{
\includegraphics[width=0.22\textwidth]{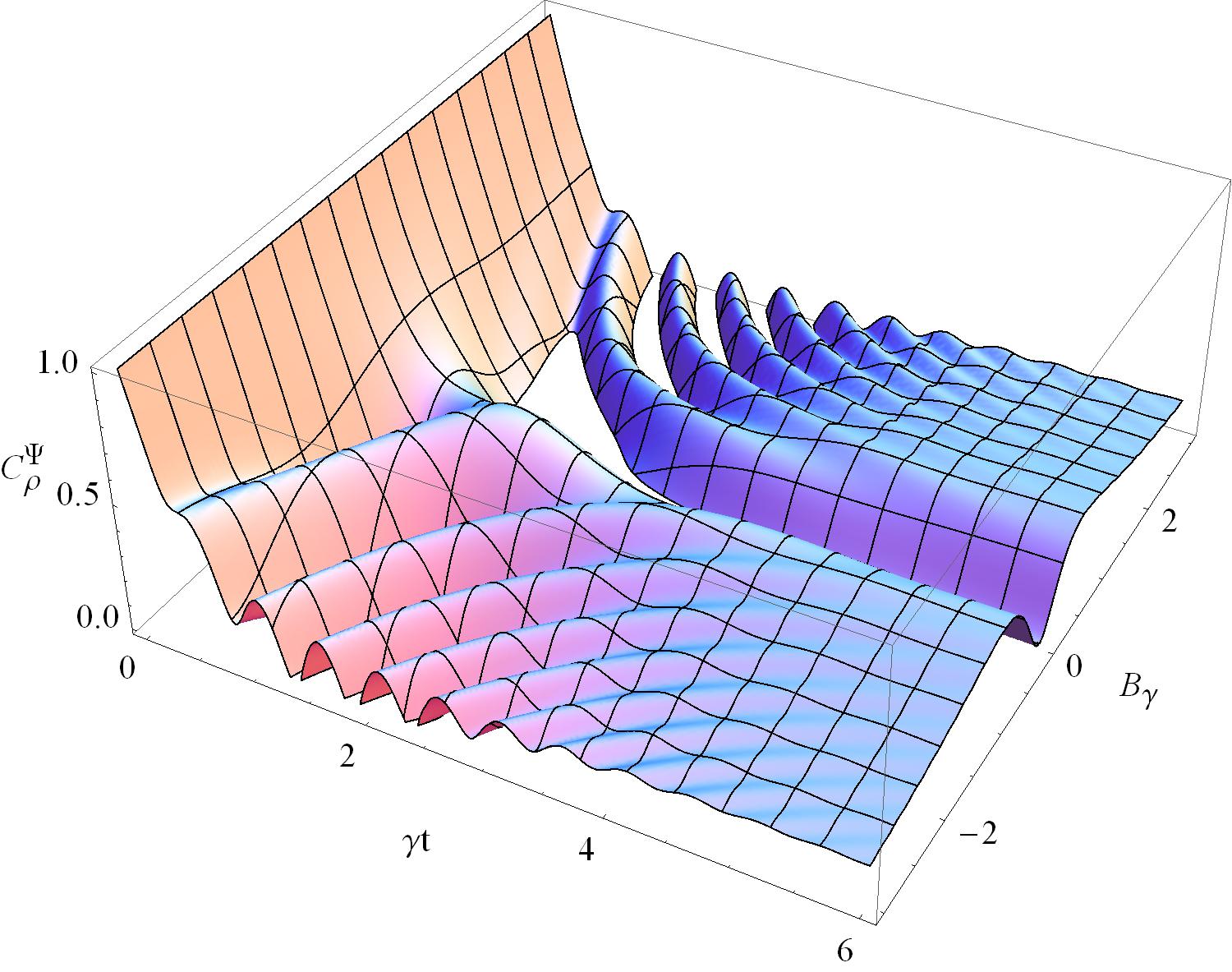}} \\
\subfigure[$a=1,b=0.$]{
\includegraphics[width=0.22\textwidth]{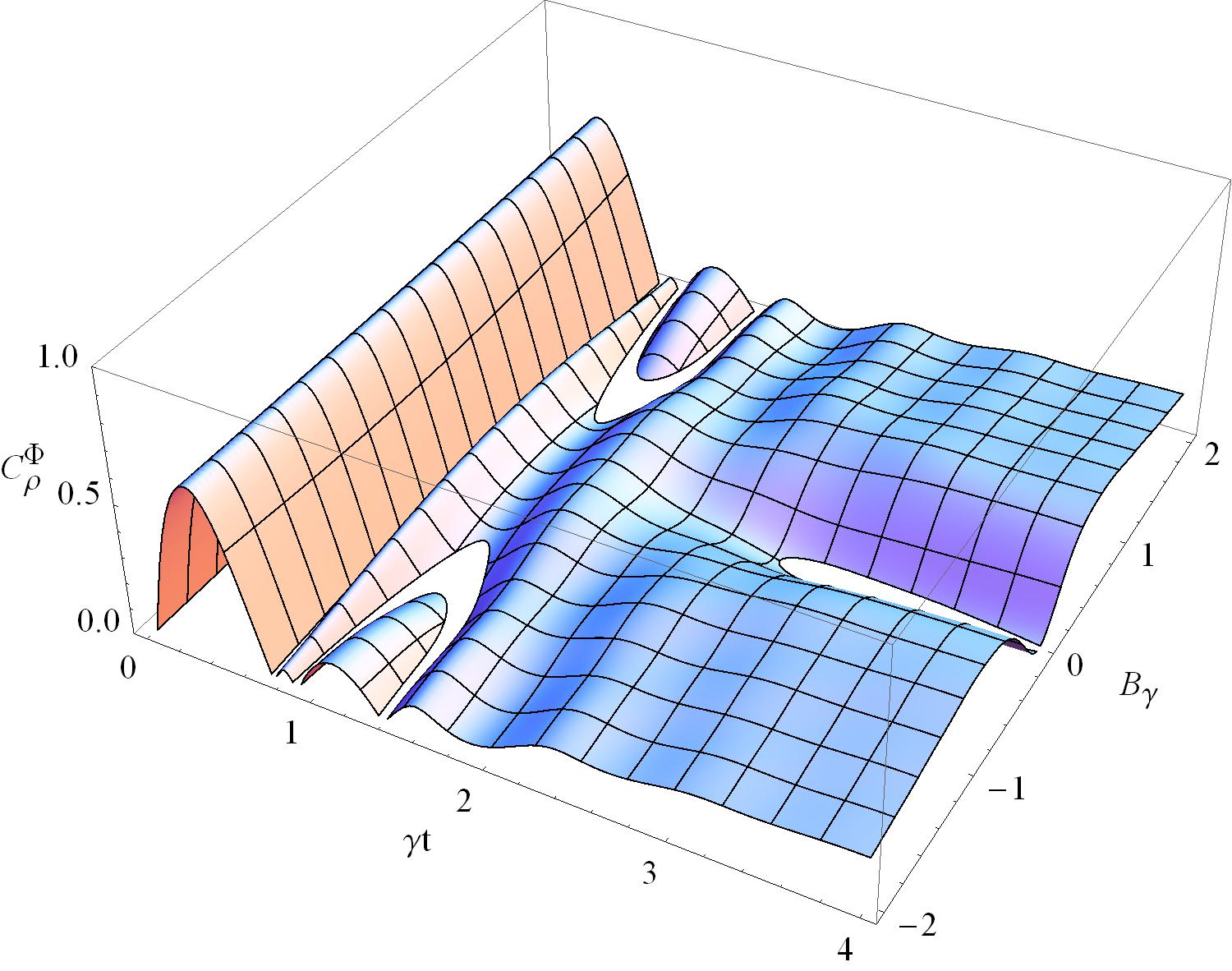}} \hspace{0.05in}
\subfigure[$a=1,b=0.$]{
\includegraphics[width=0.22\textwidth]{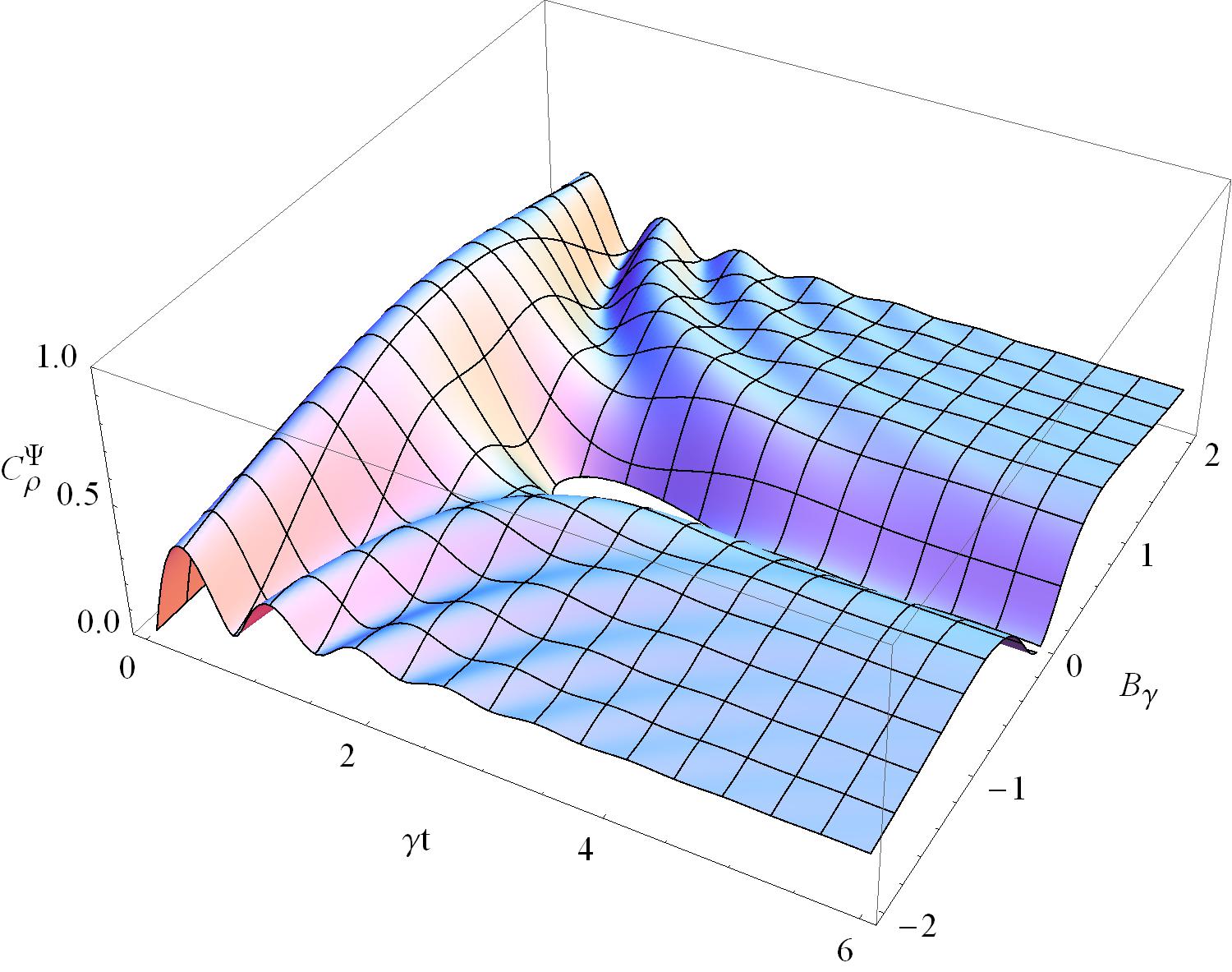}}
\caption{ $C_{\rho}\left(
t\right)  $ versus $\gamma t$ and $B_{\gamma}$ with $J\Delta_{\gamma}=1,n=0$.}
\label{fig2}
\end{figure}

\begin{figure}[H]
\setcounter{subfigure}{0} \centering
\subfigure[$r=1.$]{
\includegraphics[width=0.22\textwidth]{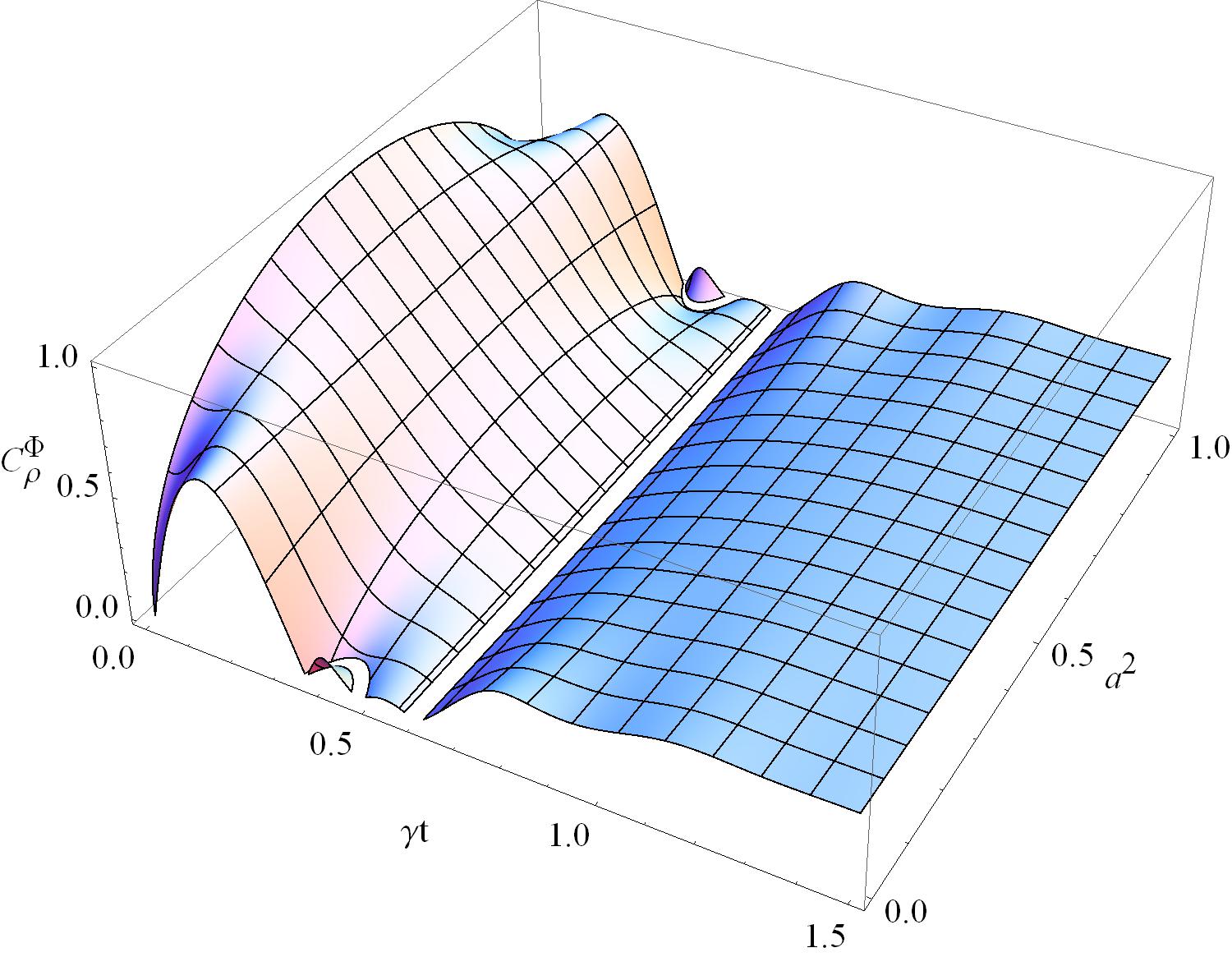}} \hspace{0.05in}
\subfigure[$r=1.$]{
\includegraphics[width=0.22\textwidth]{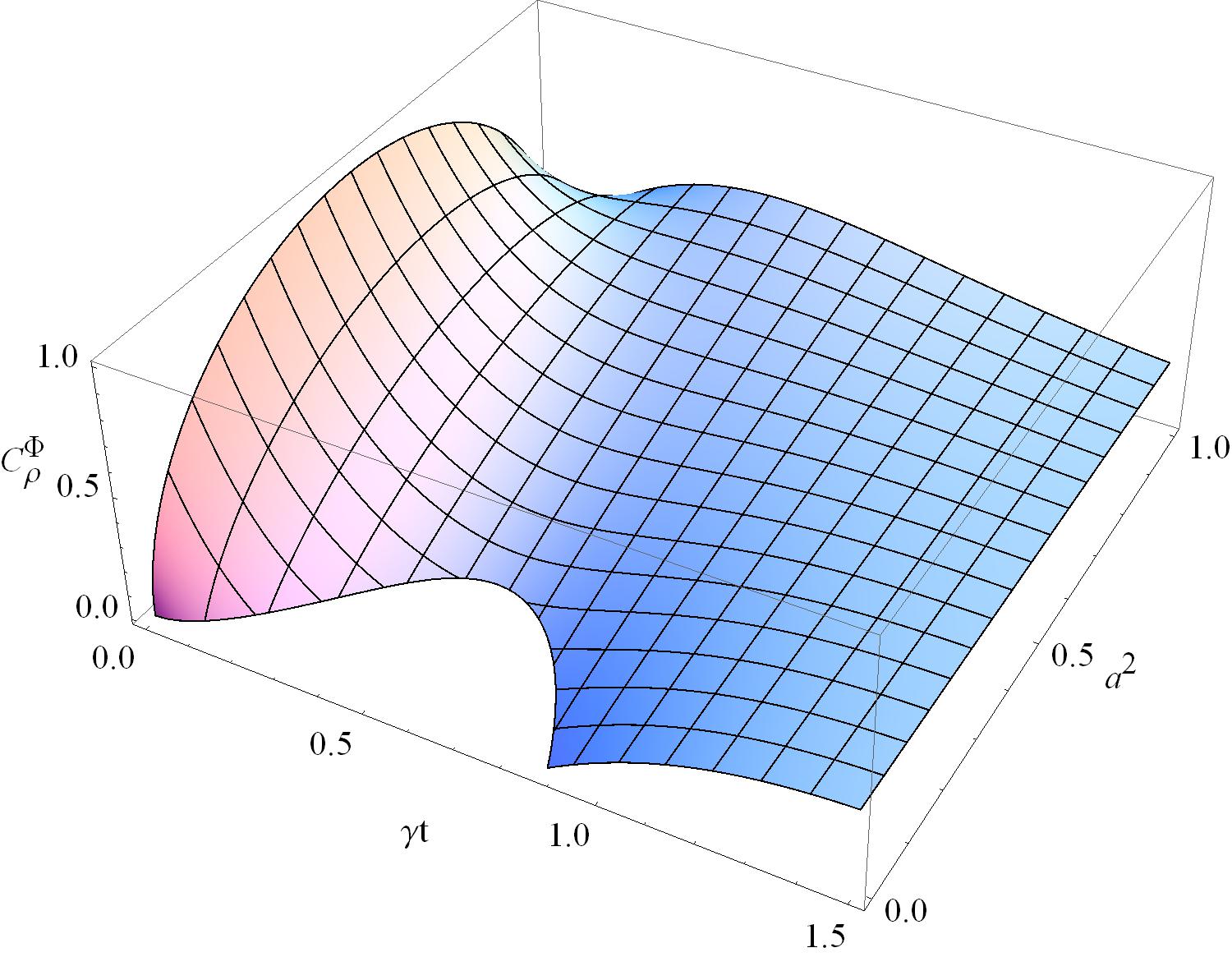}} \\
\subfigure[$r=1/2.$]{
\includegraphics[width=0.22\textwidth]{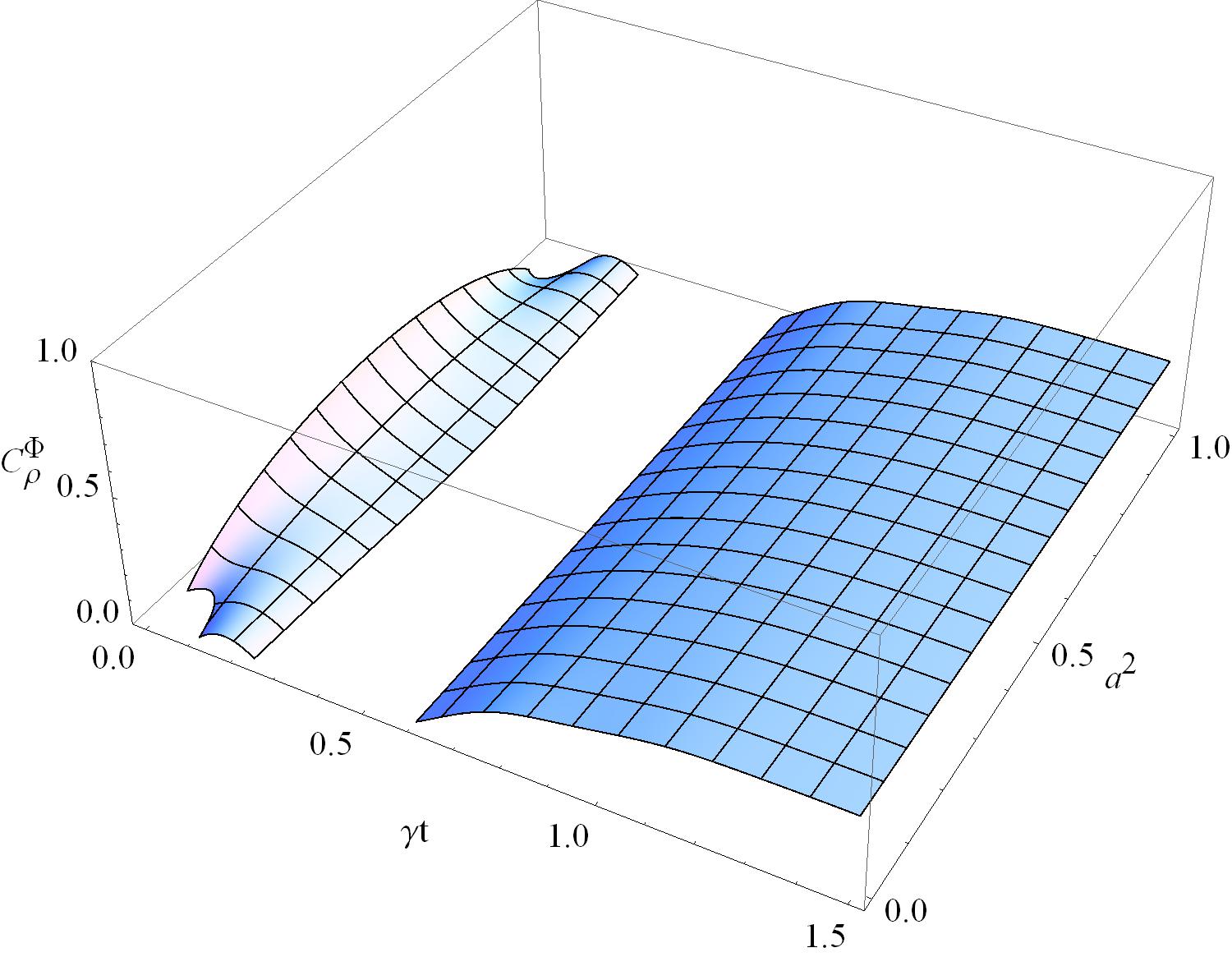}} \hspace{0.05in}
\subfigure[$r=1/2.$]{
\includegraphics[width=0.22\textwidth]{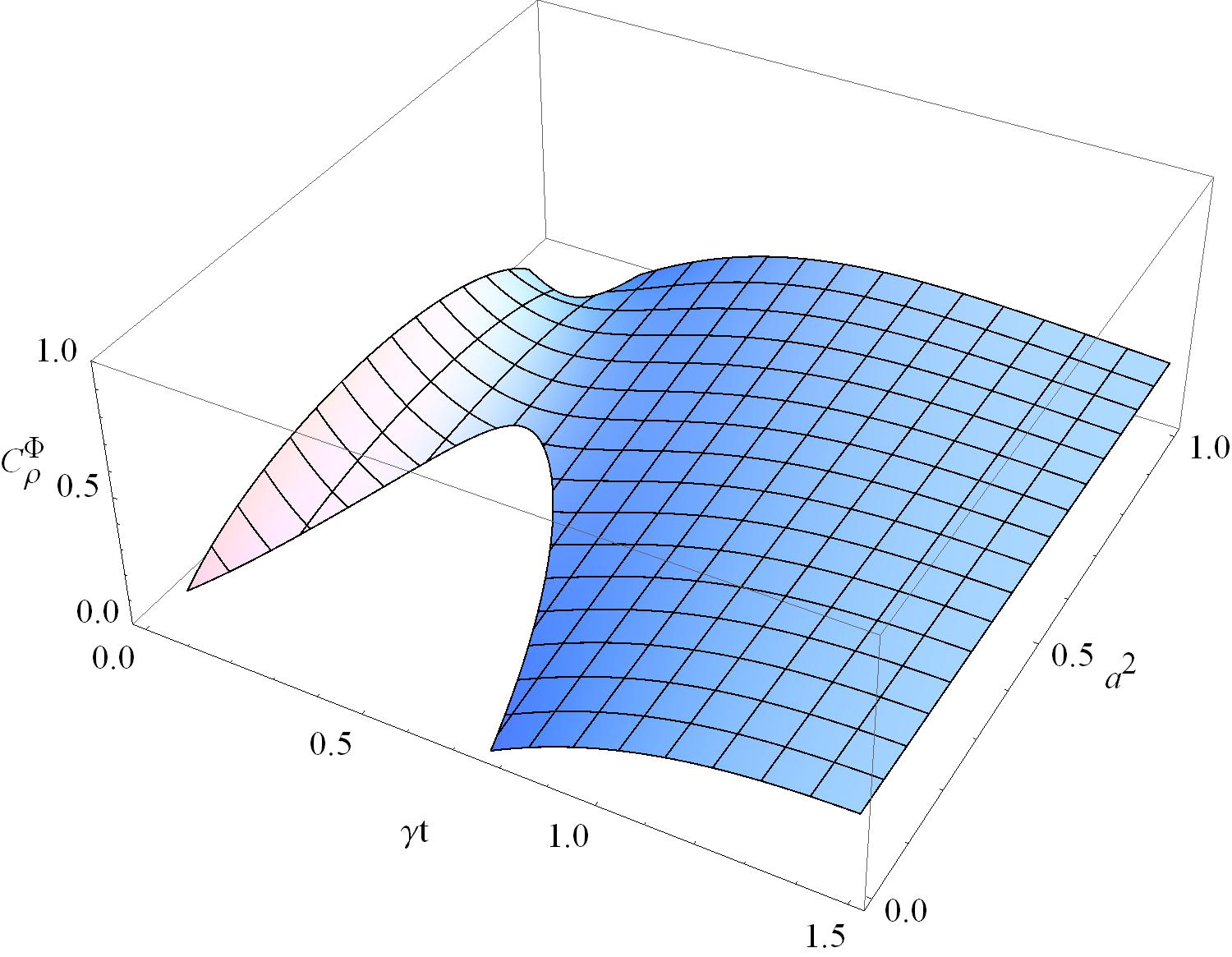}}
\caption{ $C_{\rho}\left(
t\right)  $ vs. $\gamma t$ and $a^{2}$ with $J\Delta_{\gamma}=1,n=0,B_{\gamma}=0$.}
\label{fig3}
\end{figure}

Fig.\ref{fig2} is plotted to investigate the effect of external field on concurrence dynamics. $C_{\rho}^{\Phi}\left(t\right)$ still consist of bipartite parts, ESD and revival of $C_{\rho}^{\Psi}\left(  t\right)$ occur periodically and the frequency show a positive correlation with $B_{\gamma}$.

and $a^{2}$ in Fig.\ref{fig3}.
To analyze the influences of initial entanglement, plot $C_{\rho}\left(t\right)$ in Fig.\ref{fig3}, which show $C_{\rho}^{\Phi}\left(  t\right)$ has a symmetry in the dynamic of concurrence with respect to the maximally entangled case $(a^{2}=1/2)$, while $C_{\rho}^{\Psi}\left(  t\right)  $ is asymmetry, with less concurrence for $(a^{2}<1/2)$. This can be attributed to an initially higher contribution of the spin-up states when $(a^{2}<1/2)$. Moreover, ESD and revival of $C_{\rho}^{\Psi}\left(  t\right)  $ occurs only for $a^{2}<a_{c}$, $a_{c}$ is a critical function of $J_{\gamma}\Delta,n,r$. Although the final state and concurrence is independent of $r$, the dynamic evolution is extremely influenced by which.

As is known to all, the temperature have a serious impact on dynamic of open
quantum system. Therefore we plot $C_{\rho}^{\Phi}\left(  t\right)  $ and
$C_{\rho}^{\Psi}\left(  t\right)  $ as functions of $\gamma t$ and $n$:
\begin{figure}[H]
	\setcounter{subfigure}{0} \centering
	\subfigure[$a=b=1/\sqrt{2}.$]{
		\includegraphics[width=0.22\textwidth]{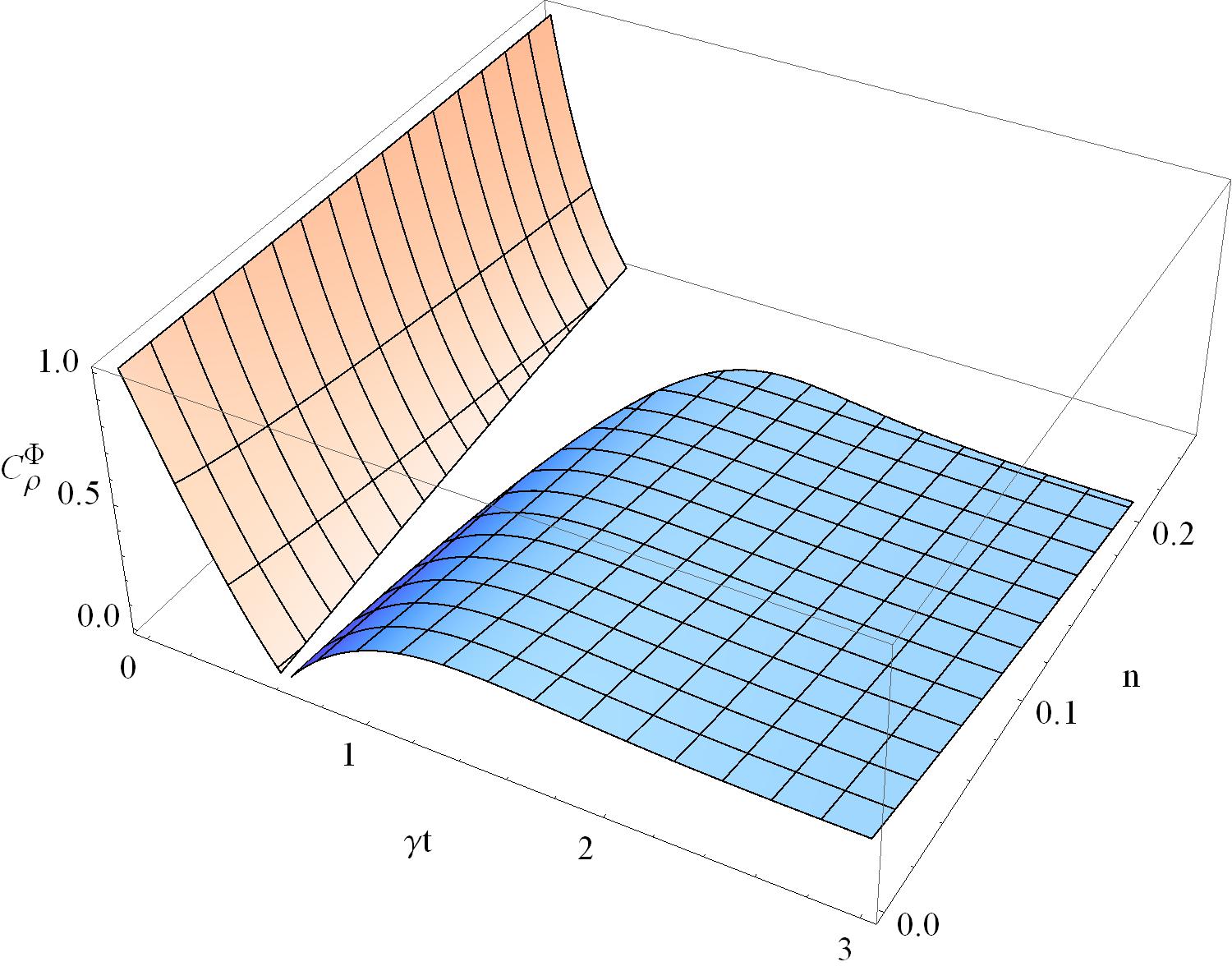}} \hspace{0.05in}
	\subfigure[$a=b=1/\sqrt{2}.$]{
		\includegraphics[width=0.22\textwidth]{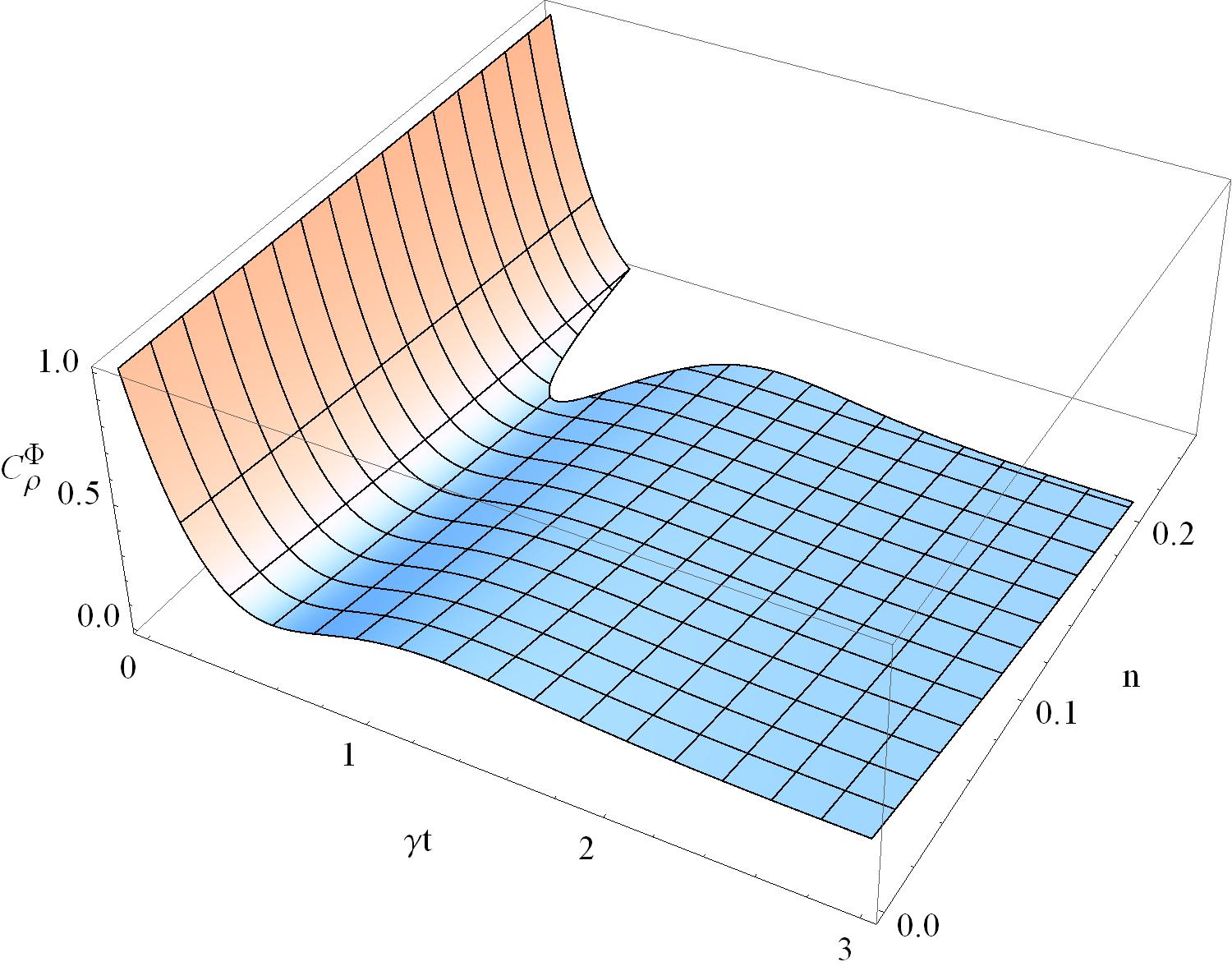}}\\
	\subfigure[$a=1,b=0.$]{
		\includegraphics[width=0.22\textwidth]{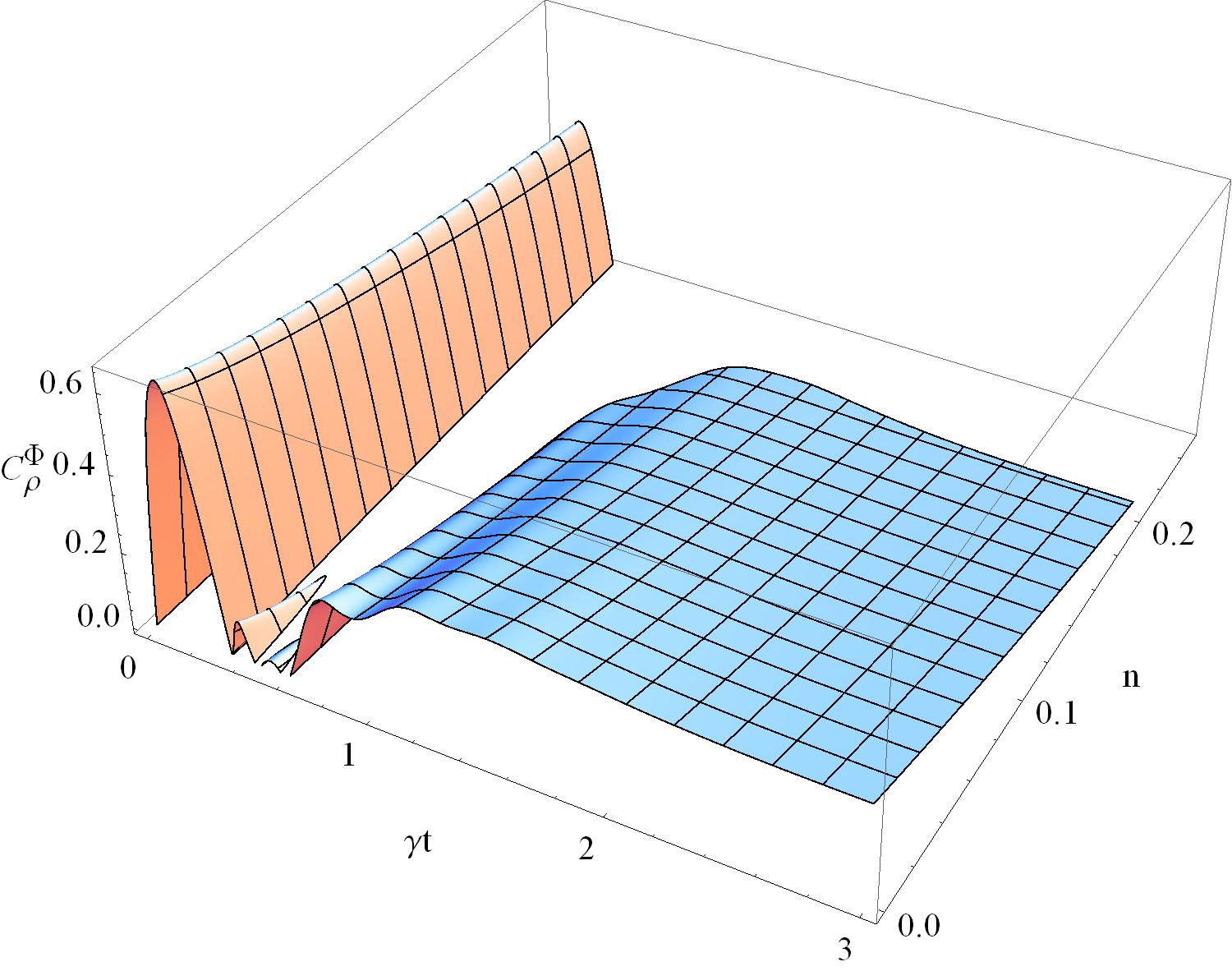}} \hspace{0.05in}
	\subfigure[$a=1,b=0.$]{
		\includegraphics[width=0.22\textwidth]{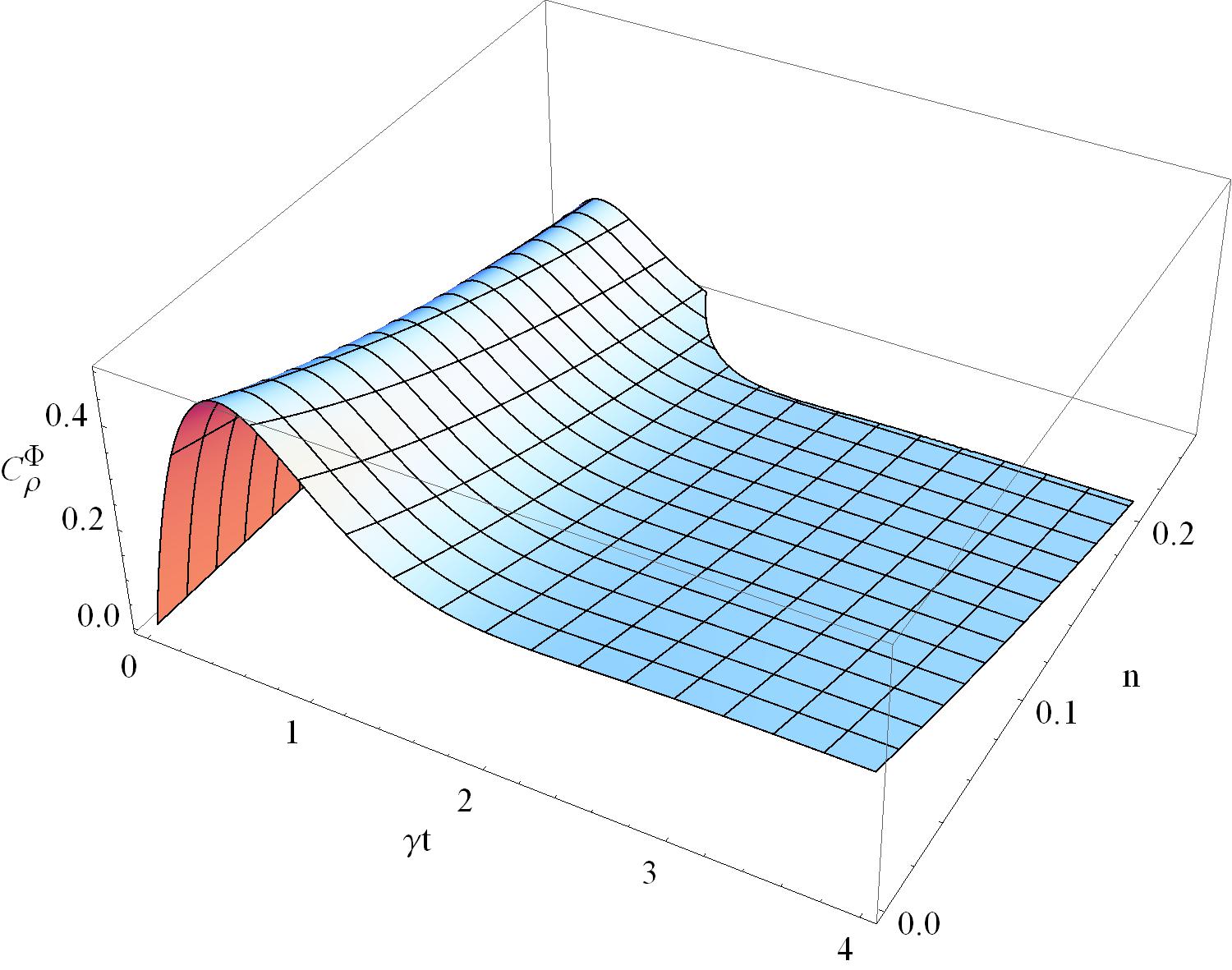}}
	\caption{ $C_{\rho}\left(
		t\right)$ versus $\gamma t$ and $n$ with $J\Delta_{\gamma}=1,r=1,B_{\gamma}=0$.}
	\label{fig4}
\end{figure}
Differ with uncoupled cases, the concurrence dynamic does not monotone decrease but fluctuate for small initial concurrence, the frequency show a positive correlation with $J$; $C_{\rho}^{\Phi}\left(  \infty\right)  $ and $C_{\rho}^{\Psi}\left(\infty\right)  $ decrease along with the growth of $n$, and $t_{ESD}$ of $C_{\rho}^{\Phi}\left(  t\right)  $ decreases whereas the $t_{revival}$ increases with the growth of $n$ respectively; Besides, both concurrence has a  maximum $(\sqrt{5}-1)/4$ for $ t\rightarrow \infty$.
This contrasts with the uncoupled $2$-qubits case, for which there is total disentanglement after a sufficient period. We shall investigate both expression and fabric of $\rho\left(  \infty\right)  ,C_{\rho}\left(  \infty\right)  $ in next section to find the causes of the difference.

\section{Asymptotic Behavior of Concurrence for $t\rightarrow\infty$}

The explicit expression of $\rho(\infty)$ and $C\left(  \infty\right)$ can be given by the stationary equation method in Sec.II, the solution is
\begin{equation}
\left\vert \rho_{t}\right\rangle =\sum_{i}C_{i}e^{\lambda_{i}t}\left\vert
\varphi_{i}\right\rangle , \label{a26}%
\end{equation}
where $\lambda_{i},\left\vert
\varphi_{i}\right\rangle$ is the eigensystem of $\mathscr{F}$,
$C_{i}$ is constant determined by initial state $\left\vert \rho
_{0}\right\rangle $.  {\bf All eigenvalues $\lambda_{i}\leq0$, so the eigenvector $\left\vert \varphi_{0}\right\rangle $ whose eigenvalue $\lambda_{0}=0$ represent the final stable state because other's coefficient $e^{\lambda_{i}t}$ damp to zero along with the increasing of $t$}
The final stable state is determined by $\mathscr{F}$ and independent of initial state, that can explain $\rho^{\Phi
}\left(  \infty\right)  =\rho^{\Psi}\left(  \infty\right)  $ and $C_{\rho}^{\Phi}\left(  \infty\right)  =C_{\rho}^{\Psi}\left(  \infty\right)  $, whose expression shall be given in appendix

Plot $C\left(
\infty\right)  $ as a function of $J_{\gamma}\Delta,B_{\gamma}$ and $n$ as follow
\onecolumn{
\begin{figure}[H]
	\setcounter{subfigure}{0} \centering
	\subfigure[$n=1/10.$]{\includegraphics[width=0.32\textwidth]{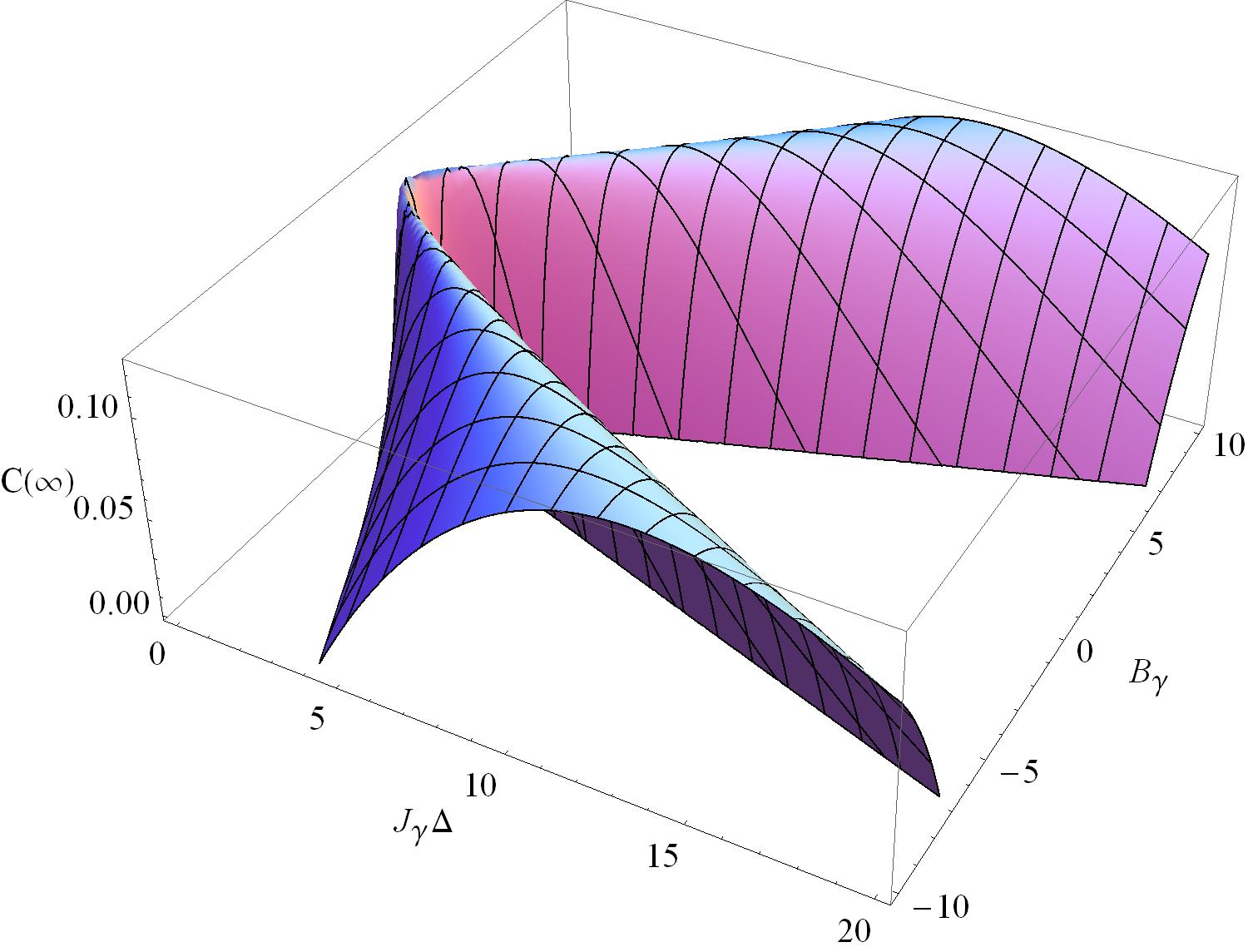}}
	\hspace{0.03in}
	\subfigure[$B_{\gamma}=1.$]{\includegraphics[width=0.32\textwidth]{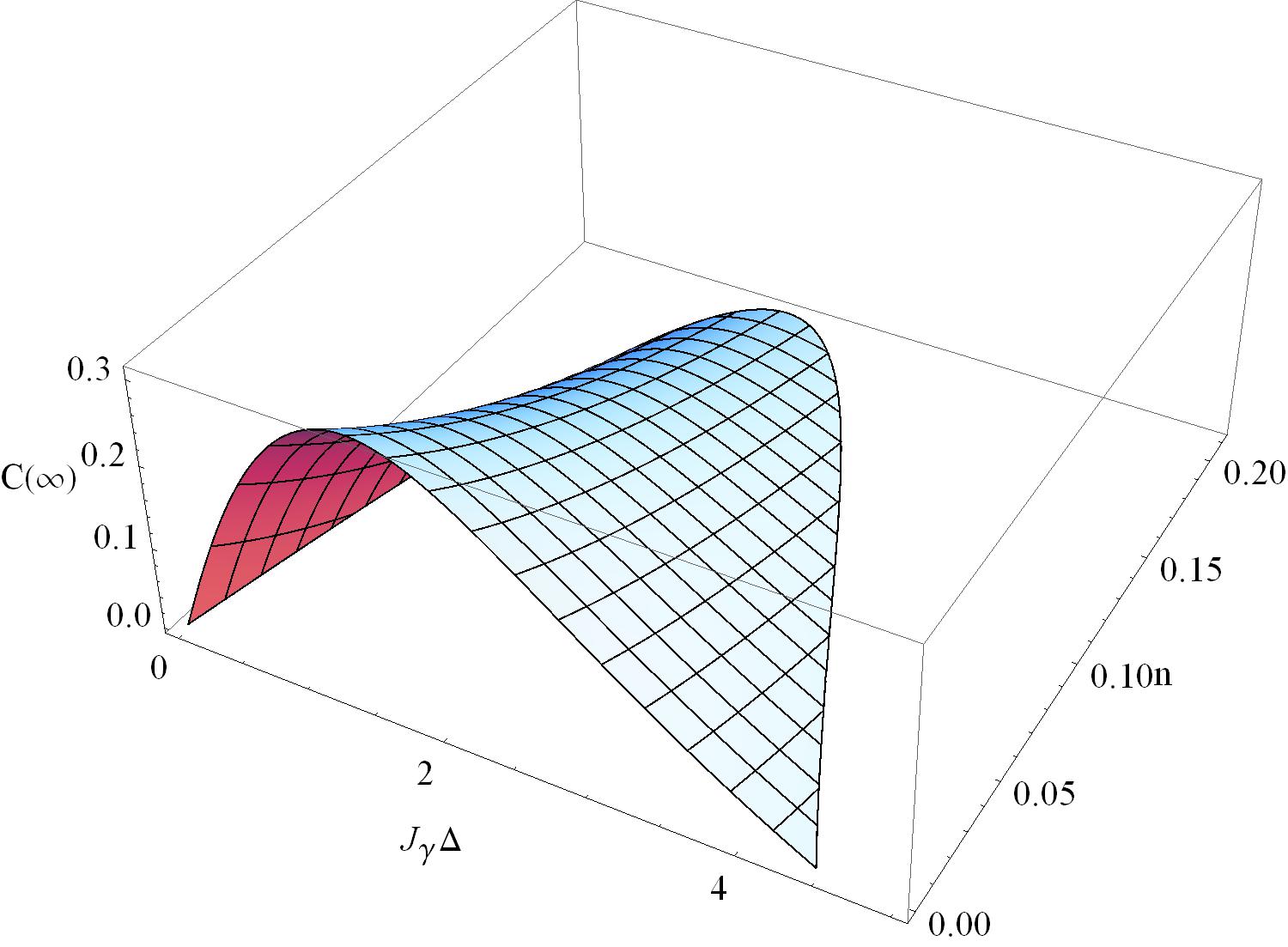}}
	\hspace{0.03in}
	\subfigure[$J_{\gamma}\Delta=1.$]{\includegraphics[width=0.32\textwidth]{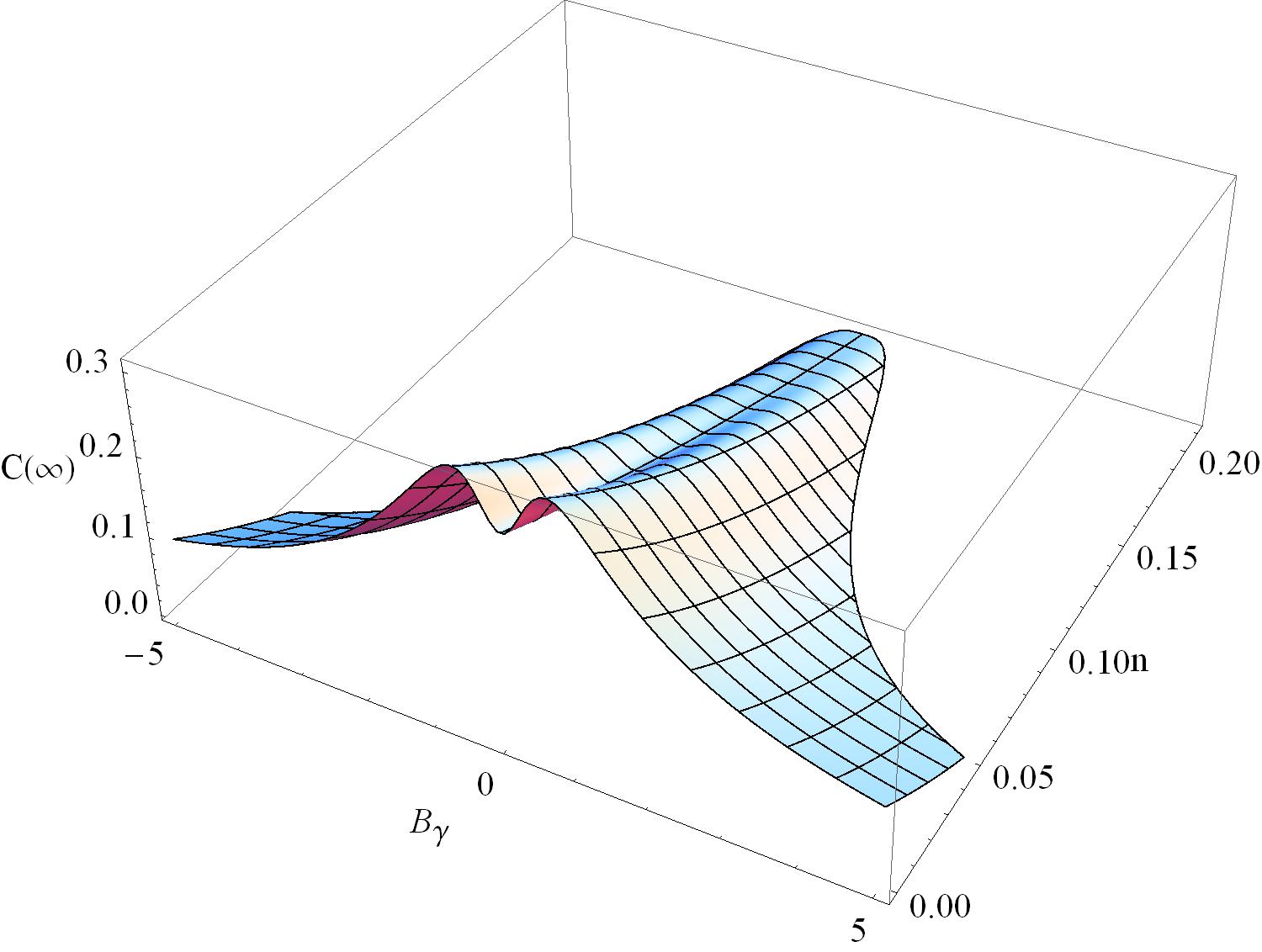}}\caption{$C\left(
		\infty\right)  \text{\ versus\ }B_{\gamma},J_{\gamma}\Delta\text{\ and\ }n.$}%
	\label{fig5}
\end{figure}}
To satisfy $C\left(\infty\right)  >0$, the necessary condition is $n<\left(  \sqrt{2}-1\right)  /2$ ,
otherwise, $C\left(  \infty\right)=0  $ any $J_{\gamma
}\Delta$ and $B_{\gamma}$; Besides, Fig.\ref{fig5} exhibit that the final concurrence take a maximum:
\begin{small}
	\begin{equation}
	C_{Max}\left(  \infty\right)  =Max[  \frac{\sqrt{5+16n(1+n)}-\left(
		8n^{2}+8n+1\right)  }{4(1+2n)^{2}},0].
	\end{equation}
\end{small}When $n=0,J_{\gamma}\Delta=\sqrt{4B^{2}+1}(\sqrt{5}-1)
/2$, the maximum is $(  \sqrt{5}-1)  /4$ just half of the golden section.

This result exposit that interacted and coupled Hamilton of $2$-qubit contribute to suppressing decoherence and disentanglement process; Noting all these occur in Markovian environment, there is no memory effect and feed back from the bath. It is the interplay between two qubits combat disentanglement eventually lead to $C_{\rho}\left(  \infty\right)  >0$, which implies the interaction of many-body system may be used to prevent the decoherence of Markovian environment.
Thus for different master equation, we may completely suppress the disentanglement process by constructing the proper $H_{spin}$ for different master equation, and finally obtain a ideal steady maximally entangled state.



\section{Conclusion}
In summary, we introduce a new Ket-Bra Enatngled State method for solving any master equation of infinite-level system. This method have a wide applicable range and can be completely calculated by computer; Besides it can convert master equation in to Schr\"{o}dinger-like equation, and use the way of Schr\"{o}dinger equation to solve master equation. Using this KBES method we solve the $2$-qubit $XYZ$ Heisenberg chain, and investigate the dynamic evolution of concurrence find that $C_{\rho}\left(  \infty\right)$ is no more alway equal to zero which is total different with uncoupled case. Then we analyze both expression and fabric of $\rho\left(\infty\right),C_{\rho}\left(  \infty\right)$ to explain the unusual phenomenon.

\end{multicols}
\appendixpage
For $\rho_{11}^{\Phi}\left( 0\right) $, The explicit solution of \kref{a16} is
\begin{equation}
\begin{array}
[c]{l}%
\rho_{11}^{\Phi}\left(  t\right)  =\frac{e^{-4t(\alpha+\beta)}}{4\left[
	(\alpha+\beta)^{2}+J^{2}\Delta^{2}\right]  }\left\{
\begin{array}
[c]{c}%
(1-r)\left(  \alpha-\beta\right)  ^{2}-4\alpha\beta r+4\beta^{2}%
e^{4t(\alpha+\beta)}+J^{2}\Delta^{2}\left(  e^{4t(\alpha+\beta)}-r\right)  \\
+2e^{2t(\alpha+\beta)}(\alpha-\beta)\left[  2\beta\cos\left(  2J\Delta
t\right)  -J\Delta\sin\left(  2J\Delta t\right)  \right]
\end{array}
\right\}  \\
\rho_{44}^{\Phi}\left(  t\right)  =\frac{e^{-4t(\alpha+\beta)}}{4\left[
	(\alpha+\beta)^{2}+J^{2}\Delta^{2}\right]  }\left\{
\begin{array}
[c]{c}%
(1-r)\left(  \alpha-\beta\right)  ^{2}-4\alpha\beta r+4\alpha^{2}%
e^{4t(\alpha+\beta)}+J^{2}\Delta^{2}\left(  e^{4t(\alpha+\beta)}-r\right)  \\
+2e^{2t(\alpha+\beta)}(\beta-\alpha)\left[  2\alpha\cos\left(  2J\Delta
t\right)  -J\Delta\sin\left(  2J\Delta t\right)  \right]
\end{array}
\right\}  \\
\rho_{22}^{\Phi}\left(  t\right)  =\frac{e^{-4t(\alpha+\beta)}}{4\left(
	(\alpha+\beta)^{2}+J^{2}\Delta^{2}\right)  }\left\{
\begin{array}
[c]{c}%
(r-1)\left(  \alpha+\beta\right)  ^{2}+4\alpha\beta\left(  e^{4t(\alpha
	+\beta)}+1\right)  +J^{2}\Delta^{2}\left(  e^{4t(\alpha+\beta)}+r\right)  +\\
2e^{2t(\alpha+\beta)}\left[  r\left(  b^{2}-a^{2}\right)  \left(
(\alpha+\beta)^{2}+J^{2}\Delta^{2}\right)  \cos\left(  2Jt\right)
+(\alpha-\beta)^{2}\cos\left(  2J\Delta t\right)  \right]
\end{array}
\right\}  \\
\rho_{22}^{\Phi}\left(  t\right)  =\frac{e^{-4t(\alpha+\beta)}}{4\left(
	(\alpha+\beta)^{2}+J^{2}\Delta^{2}\right)  }\left\{
\begin{array}
[c]{c}%
(r-1)\left(  \alpha+\beta\right)  ^{2}+4\alpha\beta\left(  e^{4t(\alpha
	+\beta)}+1\right)  +J^{2}\Delta^{2}\left(  e^{4t(\alpha+\beta)}+r\right)  +\\
2e^{2t(\alpha+\beta)}\left[  r\left(  b^{2}-a^{2}\right)  \left(
(\alpha+\beta)^{2}+J^{2}\Delta^{2}\right)  \cos\left(  2Jt\right)
+(\alpha-\beta)^{2}\cos\left(  2J\Delta t\right)  \right]
\end{array}
\right\}  \\
\rho_{33}^{\Phi}\left(  t\right)  =\frac{e^{-4t(\alpha+\beta)}}{4\left(
	(\alpha+\beta)^{2}+J^{2}\Delta^{2}\right)  }\left\{
\begin{array}
[c]{c}%
(r-1)\left(  \alpha+\beta\right)  ^{2}+4\alpha\beta\left(  e^{4t(\alpha
	+\beta)}+1\right)  +J^{2}\Delta^{2}\left(  e^{4t(\alpha+\beta)}+r\right)  +\\
2e^{2t(\alpha+\beta)}\left[  r\left(  a^{2}-b^{2}\right)  \left(
(\alpha+\beta)^{2}+J^{2}\Delta^{2}\right)  \cos\left(  2Jt\right)
+(\alpha-\beta)^{2}\cos\left(  2J\Delta t\right)  \right]
\end{array}
\right\}  \\
\rho_{23}^{\Phi}\left(  t\right)  =re^{-2t(\alpha+\beta)}\left[
ab+i\sin\left(  2Jt\right)  \left(  b^{2}-a^{2}\right)  /2\right]  \\
\rho_{14}^{\Phi}\left(  t\right)  =\frac{i(\alpha-\beta)\left\{  \left[
	J\Delta\cos\left(  2J\Delta t\right)  +(\alpha+\beta)\sin\left(
	2Jt\Delta\right)  \right]  e^{-2t(\alpha+\beta)}-J\Delta\right\}  }{2\left(
	(\alpha+\beta)^{2}+J^{2}\Delta^{2}\right)  }
\end{array}
\label{s1}
\end{equation}
While for $\rho_{11}^{\psi}\left( 0\right) $
\begin{equation}
\begin{array}
[c]{l}%
\rho_{11}^{\Psi}\left(  t\right)  =\frac{e^{-4t(\alpha+\beta)}}{4\left[
	(\alpha+\beta)^{2}+J^{2}\Delta^{2}\right]  }\left\{
\begin{array}
[c]{c}%
\left(  1-r\right)  \left(  \alpha-\beta\right)  ^{2}+4\beta^{2}%
e^{4t(\alpha+\beta)}+4r\left(  a^{2}\beta^{2}+b^{2}\alpha^{2}\right)
+J^{2}\Delta^{2}\left(  e^{4t(\alpha+\beta)}+r\right)  \\
+2\cos\left(  2J\Delta t\right)  e^{2t(\alpha+\beta)}\left[  2\beta\left(
\alpha-\beta\right)  +r\left(  b^{2}-a^{2}\right)  \left(  2\alpha\beta
+2\beta^{2}+J^{2}\Delta^{2}\right)  \right]  \\
+2J\Delta\sin\left(  2J\Delta t\right)  e^{2t(\alpha+\beta)}\left(
\beta-\alpha\right)  \left(  1+rb^{2}-ra^{2}\right)
\end{array}
\right\}  \\
\rho_{44}^{\Psi}\left(  t\right)  =\frac{e^{-4t(\alpha+\beta)}}{4\left[
	(\alpha+\beta)^{2}+J^{2}\Delta^{2}\right]  }\left\{
\begin{array}
[c]{c}%
\left(  1-r\right)  \left(  \alpha-\beta\right)  ^{2}+4\alpha^{2}%
e^{4t(\alpha+\beta)}+4r\left(  a^{2}\beta^{2}+b^{2}\alpha^{2}\right)
+J^{2}\Delta^{2}\left(  e^{4t(\alpha+\beta)}+r\right)  \\
+2\cos\left(  2J\Delta t\right)  e^{2t(\alpha+\beta)}\left[  2\alpha\left(
\beta-\alpha\right)  +r\left(  a^{2}-b^{2}\right)  \left(  2\alpha^{2}%
+2\alpha\beta+J^{2}\Delta^{2}\right)  \right]  \\
+2J\Delta\sin\left(  2J\Delta t\right)  e^{2t(\alpha+\beta)}(\alpha
-\beta)\left(  1+ra^{2}-rb^{2}\right)
\end{array}
\right\}  \\
\rho_{14}^{\Psi}\left(  t\right)  =\frac{ie^{-2t(\alpha+\beta)}}{2\left[
	(\alpha+\beta)^{2}+J^{2}\Delta^{2}\right]  }\left\{
\begin{array}
[c]{c}%
J\Delta(\alpha-\beta)\left[  \cos\left(  2J\Delta t\right)  -e^{2t(\alpha
	+\beta)}\right]  +2abr\left[  (\alpha+\beta)^{2}+J^{2}\Delta^{2}\right]  \\
+\left[  \alpha^{2}-\beta^{2}+r\left(  b^{2}-a^{2}\right)  \left(  \left(
\alpha+\beta\right)  ^{2}+J^{2}\Delta^{2}\right)  \right]  \sin\left(
2J\Delta t\right)
\end{array}
\right\}  \\
\rho_{22}^{\Psi}\left(  t\right)  =\rho_{33}^{\Psi}\left(  t\right)
=\frac{e^{-4t(\alpha+\beta)}}{4\left(  (\alpha+\beta)^{2}+J^{2}\Delta
	^{2}\right)  }\left\{
\begin{array}
[c]{c}%
\left(  r-1\right)  \left(  \alpha-\beta\right)  ^{2}+J^{2}\Delta^{2}\left(
e^{4t(\alpha+\beta)}-r\right)  -4r\left(  a^{2}\beta^{2}+b^{2}\alpha
^{2}\right)  \\
+4\alpha\beta e^{4t(\alpha+\beta)}+2rJ\Delta\sin\left(  2J\Delta t\right)
e^{2t(\alpha+\beta)}\left(  \alpha-\beta\right)  \left(  b^{2}-a^{2}\right)
\\
+2\cos\left(  2J\Delta t\right)  e^{2t(\alpha+\beta)}\left[  \left(
\alpha-\beta\right)  ^{2}+r\left(  b^{2}-a^{2}\right)  (\alpha^{2}-\beta
^{2})\right]
\end{array}
\right\}\label{s2}
\end{array}
\end{equation}
The explicit expression of $C_{\rho}^{\Phi
}\left(  t\right) ,C_{\rho}^{\Psi}\left(  t\right)  $ is
\begin{align}
C_{\rho}^{\Phi}\left(  t\right)   &  =Max\left[  K_{1}^{\Phi}\left(  t\right)
,K_{2}^{\Phi}\left(  t\right)  ,0\right]  ,\nonumber\\
C_{\rho}^{\Psi}\left(  t\right)   &  =Max\left[  K_{1}^{\Psi}\left(  t\right)
,0\right]  , \label{a23}%
\end{align}
where
\begin{align}
K_{1}^{\Phi}\left(  t\right)   &  =e^{-2\left(  \alpha+\beta\right)  t}%
\sqrt{4a^{2}b^{2}+\left(  a^{2}-b^{2}\right)  ^{2}\sin^{2}\left(  2Jt\right)
}-D^{-1}e^{-2t(\alpha+\beta)}\sqrt{G_{1}\left(  \alpha,\beta\right)
G_{1}\left(  \beta,\alpha\right)  },\nonumber\\
K_{2}^{\Phi}\left(  t\right)   &  =D^{-1}e^{-2t\left(  \alpha+\beta\right)
}\left\{  \left\vert G_{2}\left(  t\right)  \right\vert -\sqrt{G_{3}\left(
a,b\right)  G_{3}\left(  b,a\right)  }\right\}  ,\nonumber\\
K_{1}^{\Psi}\left(  t\right)   &  =D^{-1}e^{-2\left(  \alpha+\beta\right)
	t}\left[  \left\vert G_{4}\left(  t\right)  \right\vert -\left\vert
G_{5}\left(  t\right)  \right\vert \right]  , \label{a24}%
\end{align}
and
\begin{align}
D  &  =\left(  \alpha+\beta\right)  ^{2}+J^{2}\Delta^{2}.\nonumber\\
G_{1}\left(  \alpha,\beta\right)   &  =\left(  \alpha-\beta\right)  \left[
2\alpha\cos\left(  2J\Delta t\right)  -J\Delta\sin\left(  2J\Delta t\right)
\right]  -\left(  2\alpha^{2}+\frac{J^{2}\Delta^{2}}{2}\right)  e^{2\left(
	\alpha+\beta\right)  t}+\left(  2\alpha\beta+\frac{J^{2}\Delta^{2}}{2}\right)
e^{-2\left(  \alpha+\beta\right)  t}.\nonumber\\
G_{2}\left(  t\right)   &  =\left(  \alpha-\beta\right)  \left[  J\Delta
\cos\left(  2J\Delta t\right)  +\left(  \alpha+\beta\right)  \sin\left(
2J\Delta t\right)  -J\Delta e^{2t(\alpha+\beta)}\right]  .\nonumber\\
G_{3}\left(  a,b\right)   &  =D\left(  a^{2}-b^{2}\right)  \cos\left(
2Jt\right)  +\cosh\left[  2\left(  \alpha+\beta\right)  t\right]  \left(
J^{2}\Delta^{2}+4\alpha\beta\right)  +\left(  \alpha-\beta\right)  ^{2}%
\cos\left(  2J\Delta t\right)  .\nonumber\\
G_{4}\left(  t\right)   &  =\left(  a-\beta\right)  J\Delta\left[  e^{2\left(
	\alpha+\beta\right)  t}-\cos\left(  2J\Delta t\right)  \right]  -2iabD+\left(
2\beta\left(  \alpha+\beta\right)  +J^{2}\Delta^{2}-2Db^{2}\right)
\sin\left(  2J\Delta t\right)  .\nonumber\\
G_{5}\left(  t\right)   &  =2\left(  b^{2}\alpha^{2}+a^{2}\beta^{2}\right)
\left(  \cos\left(  2J\Delta t\right)  -e^{-2\left(  \alpha+\beta\right)
	t}\right)  +2\alpha\beta\left(  e^{2\left(  \alpha+\beta\right)  t}%
-\cos\left(  2J\Delta t\right)  \right) \nonumber\\
&  +J^{2}\Delta^{2}\sinh\left[  2\left(  \alpha+\beta\right)  t\right]
+\frac{1}{2}\left(  b^{2}-a^{2}\right)  J\Delta e^{-2\left(  \alpha
	+\beta\right)  t}\sin\left(  2J\Delta t\right)  . \label{a27}%
\end{align}

The expression of $\rho\left( t\right)$ when $ t\rightarrow\infty$ is
		\begin{equation}
		\rho\left(  t\rightarrow\infty\right)  =%
		\begin{bmatrix}%
		\begin{smallmatrix}
		\frac{J^{2}\Delta^{2}(\alpha+\beta)^{2}+4\beta^{2}\left(  4B^{2}+(\alpha
			+\beta)^{2}\right)  }{4(\alpha+\beta)^{2}\left(  4B^{2}+J^{2}\Delta
			^{2}+(\alpha+\beta)^{2}\right)  } & 0 & 0 & \frac{-iJ\Delta(\alpha
			-\beta)(-2iB+\alpha+\beta)}{2(\alpha+\beta)\left(  4B^{2}+J^{2}\Delta
			^{2}+(\alpha+\beta)^{2}\right)  }\\
		0 & \frac{J^{2}\Delta^{2}(\alpha+\beta)^{2}+4\alpha\beta\left(  4B^{2}%
			+(\alpha+\beta)^{2}\right)  }{4(\alpha+\beta)^{2}\left(  4B^{2}+J^{2}%
			\Delta^{2}+(\alpha+\beta)^{2}\right)  } & 0 & 0\\
		0 & 0 & \frac{J^{2}\Delta^{2}(\alpha+\beta)^{2}+4\alpha\beta\left(
			4B^{2}+(\alpha+\beta)^{2}\right)  }{4(\alpha+\beta)^{2}\left(  4B^{2}%
			+J^{2}\Delta^{2}+(\alpha+\beta)^{2}\right)  } & 0\\
		\frac{iJ\Delta(\alpha-\beta)(2iB+\alpha+\beta)}{2(\alpha+\beta)\left(
			4B^{2}+J^{2}\Delta^{2}+(\alpha+\beta)^{2}\right)  } & 0 & 0 & \frac
		{16B^{2}\alpha^{2}+\left(  J^{2}\Delta^{2}+4\alpha^{2}\right)  (\alpha
			+\beta)^{2}}{4(\alpha+\beta)^{2}\left(  4B^{2}+J^{2}\Delta^{2}+(\alpha
			+\beta)^{2}\right)  }
		\end{smallmatrix}
		\end{bmatrix}.
		\label{a28}%
		\end{equation}
	The corresponding concurrence is $C\left(  \infty\right)  =Max\left[
	K_{M},0\right]  $, where%
	\begin{equation}
	K_{M}=\frac{2J\Delta\left(  \alpha^{2}-\beta^{2}\right)  \sqrt{4B^{2}
			+(\alpha+\beta)^{2}}-4\alpha\beta\left(  4B^{2}+(\alpha+\beta)^{2}\right)
		-J^{2}\Delta^{2}(\alpha+\beta)^{2}}{2(\alpha+\beta)^{2}\left(  4B^{2}
		+J^{2}\Delta^{2}+(\alpha+\beta)^{2}\right)  } \label{a29}%
	\end{equation}
\end{document}